\documentclass[a4paper,11pt]{article}
\pdfoutput=1 % if you are submitting a pdf latex (i.e. if you have
\usepackage[utf8]{inputenc}
\usepackage{jheppub} % for details on the use of the package, please
\usepackage{slashed}                     % see the JHEP-author-manual
\usepackage{braket}
\usepackage{mathtools}
\usepackage[T1]{fontenc} % if needed
\graphicspath{{img/}} % set images path

\title{\boldmath $2$-R\'enyi CCNR Negativity of Compact Boson for multiple disjoint intervals}

 \author{Himanshu Gaur}
 \affiliation{Department of Physics, Indian Institute of Technology Bombay, Powai, Mumbai, Maharashtra 400076 India}

\emailAdd{194123018@iitb.ac.in}

\abstract{We investigate mixed-state bipartite entanglement between multiple disjoint intervals using the computable cross-norm criterion (CCNR). We consider entanglement between a single interval and the union of remaining disjoint intervals, and compute $2$-R\'enyi CCNR negativity for a $2$d massless compact boson. The expression for $2$-R\'enyi CCNR negativity is given in terms of cross-ratios and Riemann period matrices of Riemann surfaces involved in the calculation. In general, the Riemann surfaces involved in the calculation of $n$-R\'enyi CCNR negativity do not possess a $Z_n$ symmetry. We also evaluate the reflected R\'enyi entropy related to the $2$-R\'enyi CCNR negativity. This reflected R\'enyi entropy is a universal quantity. We extend these calculations to the $2$d massless Dirac fermions as well. Finally, the analytical results are checked against the numerical evaluations in the tight-binding model and are found to be in good agreement.}

\makeatletter
\gdef\@fpheader{}
\makeatother
\begin{document} 
\maketitle
\flushbottom

\section{Introduction} \label{introduction}
Entanglement has been a driving force behind some of the recent developments in many frontiers of physics. Specifically entanglement has provided key insights in gravity \cite{Ryu:2006bv, Nishioka:2009un, Solodukhin:2011gn}, quantum computation \cite{Nielsen_Chuang_2010}, and quantum many-body systems \cite{Amico:2007ag, Osterloh:2002sym, Vidal:2002rm, Calabrese:2004eu}.

Entanglement and R{\'e}nyi entropies provide complete characterisation of entanglement for pure states in bipartite systems. However, these entropy-based entanglement measures fail when the system is in a mixed state, as they fail to distinguish between the quantum and classical correlations. The study of mixed-state entanglement is in general a challenging problem. To address this issue, entanglement measures based on separability criteria, such as negativity measures, are introduced to determine whether a state is entangled or separable. Among the separability criteria, entanglement measures based on positive partial transpose (PPT) criterion \cite{Peres:1996dw, Horodecki:1996nc} have been extensively studied in quantum many-body systems and quantum field theories. For a bipartite state $\rho$ shared between subsystems $A$ and $B$, PPT criterion states that the state is entangled if the trace norm of the partially transposed density matrix (say over subsystem $B$) exceeds unity, i.e. $||\rho^{T_B}||_1>1.$ Recently, negativity measures based on computable cross norm (CCNR) criterion \cite{Rudolph:2005tpa, Rudolph:2002qos, Chen:2003slc}, which is discussed in the next section, have also been studied in the quantum many-body systems and field theories \cite{Yin:2022toc}. 

Since these separability criteria can be inconclusive in detecting entanglement for general settings, these two criteria serve as complementary tools in the analysis of mixed-state entanglement, especially in quantum many-body and quantum field theories, where they are computationally accessible. Further, it is not known whether one criterion has a universal advantage over the other in discerning entanglement \cite{Rudolph:2005tpa}. These two criteria also differ in a key property that while PPT negativity is non-increasing on average under local operations and classical communication (LOCC) \cite{Vidal:2002zz, Plenio:2005cwa}, the same is not true for CCNR negativity \cite{Rudolph:2002qos, Berthiere:2023gkx}. It is also noteworthy that mathematically the CCNR negativity is also closely related to operator entanglement \cite{Berthiere:2023gkx}, as well as reflected entropies \cite{Milekhin:2022zsy}, first introduced in the context of holography \cite{Dutta:2019gen}.

Entanglement has proven to be an indispensable tool in the study of critical systems \cite{Amico:2007ag}, owing to its ability to capture the scaling of quantum correlations near the quantum phase transitions \cite{Holzhey:1994we, Calabrese:2004eu,Osterloh:2002sym, Vidal:2002rm}. This has motivated extensive research investigating entanglement properties in $2$d conformal field theories (CFTs). Entanglement studies in CFTs have been made in the context of critical systems \cite{Calabrese:2004eu, Calabrese:2009qy, Fradkin:2006mb, Cardy:2016fqc, Alcaraz:2011tn, Hsu:2008af, Calabrese:2014yza, Shapourian:2016cqu}, non-equilibrium dynamics \cite{Calabrese:2005in, Calabrese:2007mtj, Cardy:2016fqc, Wen:2015qwa, Coser:2014gsa, Hoogeveen:2014bqa}, integrable models \cite{Cardy:2007mb, Doyon:2008vu, Castro-Alvaredo:2009yqb, Bianchini:2015uea, Blondeau-Fournier:2015yoa, Castro-Alvaredo:2018dja, Castro-Alvaredo:2008fni}, and systems with boundaries or defects \cite{zhou2006entanglement, Cornfeld:2017tkz, Estienne:2023ekf, Capizzi:2022xdt, Capizzi:2022uni, Gutperle:2017enx, Ohmori:2014eia, Rogerson:2022yim, Capizzi:2022igy}. Recently, there has been a lot of interest in studying symmetry-resolved entanglement in CFTs \cite{Castro-Alvaredo:2024azg, Goldstein:2017bua, Cornfeld:2018wbg, Xavier:2018kqb, Murciano:2021djk, Jones:2022tgp, Murciano:2020vgh, Foligno:2022ltu, Capizzi:2021kys, Horvath:2021fks, Parez:2020vsp, DiGiulio:2022jjd, Kusuki:2023bsp, Calabrese:2021wvi, zhao2021symmetry, weisenberger2021symmetry, zhao2022charged} as well.

The ground-state entanglement and R{\'e}nyi entropies of a single interval in a $2$d conformal field theory (CFT) reveal the central charge of the CFT \cite{Holzhey:1994we}. When the subsystem consists of an arbitrary number of disjoint intervals, entanglement measures become significantly richer, encoding the full local operator content of the CFT and thereby completely identifying the CFT \cite{Furukawa:2008uk,Calabrese:2009ez, Calabrese:2010he, Coser:2013qda, Headrick:2012fk, Coser:2015dvp, Alba:2011fu, Rajabpour:2011pt, Ruggiero:2018hyl, Chen:2013kpa, Li:2016pwu, Ares:2022gjb, Gaur:2023yru}. In practice, the replica trick is employed to compute R{\'e}nyi entropy $S_n$ for integer values of $n$. This method involves evaluating partition functions on Riemann surfaces whose genus depends on the R{\'e}nyi index $n$, thus making R{\'e}nyi entropies sensitive to the local operator content. While the entanglement entropy is formally obtained by taking the limit $n\to 1$, this analytic continuation remains a subtle and largely open problem in general, due to the non-trivial dependence of the Riemann surface's genus on $n$.

Another interesting setting is to consider the ground state entanglement among the disjoint intervals themselves. Since the reduced state of such a subsystem is mixed, entanglement negativity measures are particularly relevant in this context. The R{\'e}nyi PPT negativity $R_n$ has been extensively studied for the case of two disjoint intervals \cite{Calabrese:2012ew, Calabrese:2012nk, Coser:2015eba, DeNobili:2015dla, Chen:2021nma, Gaur:2022sjf}. These calculations also rely on the replica trick and Riemann surfaces analogous to the ones in entropy computations arise, thus also making R{\'e}nyi negativities sensitive to the local operator content of the CFT. Consequently, the same technical challenge of taking the analytic continuation in $n$ to extract the logarithmic negativity remains an open problem.

Recently, Rényi CCNR negativities $\mathcal{E}_n$ have also been employed to study mixed-state entanglement of two-disjoint intervals \cite{Yin:2022toc}. A key finding is that the associated Riemann surfaces arising in these computations are always tori, independent of the Rényi index $n$. This simplification enables a well-defined analytic continuation in $n$, allowing the evaluation of CCNR logarithmic negativity. This computational difference highlights that the CCNR negativities characterise entanglement in a fundamentally distinct manner from the PPT R{\'e}nyi negativities and even R{\'e}nyi entropies in 2$d$ CFTs. CCNR negativity has also been studied in topological systems \cite{Yin:2023jad}, and its symmetry resolution has been studied in quantum-many-body systems and CFTs \cite{Berthiere:2023gkx, Bruno:2023tez}.

In this work, we extend the study of R{\'e}nyi CCNR negativities to the case of multiple disjoint intervals. Given the distinctive structure exhibited by CCNR negativities for two disjoint intervals in $2$d CFTs, it is interesting to explore their behaviour in more complex configurations involving multiple intervals. The pure-state entanglement of multiple disjoint intervals has already been studied through R{\'e}nyi entropies \cite{Coser:2013qda}. As we will see later, the corresponding Riemann surfaces that arise in the computation of R{\'e}nyi CCNR negativities in such settings once again become sensitive to the R{\'e}nyi index $n$. However, these surfaces have distinct characteristics from those appearing in the computation of R{\'e}nyi entropies and R{\'e}nyi PPT negativities. In particular, the latter possesses a $Z_n$ symmetry, which is typically absent in the CCNR case. 

\begin{figure}
\centering % \begin{center}/\end{center} takes some additional vertical space
%\includegraphics[width=.5\textwidth]{tikz1.pdf}
% "\includegraphics" is very powerful; the graphics package is already loaded
\includegraphics[width=0.8\textwidth]{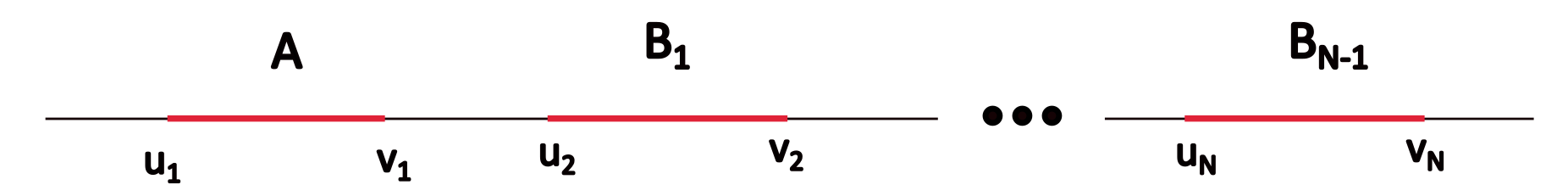}
\caption{\label{fig:i} Multiple Disjoint Intervals. The interval A is $(u_1,v_1)$ and the intervals $B_i$ are $(u_{i-1},v_{i-1})$}
\end{figure}

In the present work, we consider entanglement between a single interval, denoted $A$, and the rest of the disjoint intervals, denoted $B_i$, see Figure \ref{fig:i} and compute the $2$-R\'enyi CCNR negativity. We hope to generalise this result to arbitrary integer values of the R\'enyi index in future work. We will focus our attention on the $2$d compact boson CFT for our study.

The organisation of this work is as follows. In Section \ref{section2}, we discuss the R\'enyi CCNR negativity and reflected entropies. In Section \ref{section3}, we review the replica trick and twist fields used in the evaluation of the R\'enyi CCNR negativities. In section \ref{section4}, we evaluate the $2$-R\'enyi CCNR negativity, and the related reflected entropy for compact boson and also extend these results to massless Dirac fermion. We also numerically check our results against the tight-binding model. Finally, in Section \ref{section5}, we conclude the present work. We also have three appendices, containing necessary calculations, background material and details of the numerical model used in the present study.    

\section{Computable Cross Norm Criterion} \label{section2}
In this section, we will discuss the R\'enyi computable cross-norm negativity and reflected entropies. We consider a bipartite system composed of subsystems $A$ and $B$ in a state $\rho$, where $\rho$ can be a mixed-density matrix. Let us denote by $\{\ket{a}\}$, and $\{\ket{b}\}$ a complete set of basis on the Hilbert spaces $\mathcal{H}_A$ and $\mathcal{H}_B$, respectively. The key idea in computable cross norm negativity measures is to introduce realignment matrix $R$, given by
\begin{equation} \label{eq2.1}
\bra{a_1}\bra{a_2}R\ket{b_1}\ket{b_2}= \bra{a_1}\bra{b_1}\rho\ket{a_2}\ket{b_2}.
\end{equation}
A direct consequence of realignment is that $R$ is not a square matrix. Consider the singular value decomposition (SVD) of $R=U\Sigma V^{\dagger}$, where $\Sigma$ is a diagonal matrix, then the computable cross norm (denoted CCNR or sometimes CCN) of $R$ is defined as $\mathrm{Tr}(\Sigma)$. It is known that CCNR is a separability criterion, that is, for separable density matrices $\rho$, the CCNR of realignment matrix $R$ is less than or equal to unity. It then follows that a CCNR of greater than unity means $\rho$ is entangled. However, just like the positive partial transpose criterion, the converse is not necessarily true \cite{Rudolph:2002qos, Chen:2003slc}. To evaluate $\mathrm{Tr}(\Sigma)$, let us consider the square matrix $RR^{\dagger}$, and define the R\'enyi CCNR negativity $\mathcal{E}_n$ as
\begin{equation} \label{eq2.2}
\mathcal{E}_n=\log\mathrm{Tr}\left(RR^{\dagger}\right)^n,
\end{equation} 
The CCNR negativity is defined as the logarithm of the computable cross norm of $R$, and is given by
\begin{equation} \label{eq2.3}
\mathcal{E}=\lim_{n\to\frac{1}{2}}\mathcal{E}_n.
\end{equation}
It is known that R\'enyi CCNR negativity is closely related to reflected entropy \cite{Milekhin:2022zsy}. Let us briefly discuss this relation here. To introduce reflected entropies, we consider the same settings as above. The idea is to introduce a purification $\ket{\Omega}$ of $\rho^m$. This is done using the Choi-Jamiolkowski isomorphism,
\begin{equation} \label{eq2.4}
\ket{\Omega_m}=\frac{1}{\sqrt{\mathrm{Tr}\rho^m}}\sum_{a_1,b_1,a_2,b_2}\bra{a_1b_1}\rho^{m/2}\ket{a_2b_2}\ket{a_1b_1}\ket{a_2b_2},
\end{equation}
where the state $\ket{\Omega_m}$ belongs to the doubled Hilbert space $\mathcal{H}_A\otimes\mathcal{H}_B\otimes\mathcal{H}_A\otimes\mathcal{H}_B$. Reflected entropies characterise entanglement between the doubled subsystems $AA$ and $BB$. The reflected reduced density matrix is 
\begin{equation} \label{eq2.5}
\rho_m(AA)=\mathrm{Tr}_{BB}\left(|{\Omega_m}\rangle \langle{\Omega_m}|\right).
\end{equation}
The reflected R\'enyi entropies are then given by
\begin{equation} \label{eq2.6}
S_{m,n}(AA)=\frac{1}{1-n}\log\mathrm{Tr}\rho_m^n(AA).
\end{equation}
We see from eq.\eqref{eq2.1}-\eqref{eq2.2} that the R\'enyi CCNR negativity is just the unnormalised $m=2$ reflected R\'enyi entropy (upto a factor of $1-n$),
\begin{equation} \label{eq2.7}
S_{2,n}(AA)=\frac{1}{1-n}\mathcal{E}_n-\frac{n}{1-n}\log\left(\mathrm{Tr}\rho_{AA}^2\right).
\end{equation} 
As we will see later, $S_{2,n}(AA)$ is a universal quantity in critical systems. Finally, let us mention that generalisations of R\'enyi CCNR negativity, denoted $(m,n)$-R\'enyi CCNR negativities, have been proposed in \cite{Berthiere:2023gkx} which satisfy relations with reflected entropies similar to eq.\eqref{eq2.7} for general values of $m$.

\section{Replica Trick and Twist Fields} \label{section3}
Let us start our discussion by considering the case of two disjoint intervals, as we will see that for this case the Riemann surface associated with the replica trick for the moments of $RR^{\dagger}$ is always a torus \cite{Yin:2022toc}. We will then end this section by setting up the calculations for the case of multiple disjoint intervals. We will also see that the associated Riemann surface in this case doesn't have a fixed genus, but its genus increases with R{\'e}nyi index $n$.

\begin{figure}
\centering % \begin{center}/\end{center} takes some additional vertical space
%\includegraphics[width=.5\textwidth]{tikz1.pdf}
% "\includegraphics" is very powerful; the graphicx package is already loaded
\includegraphics[width=0.5\textwidth]{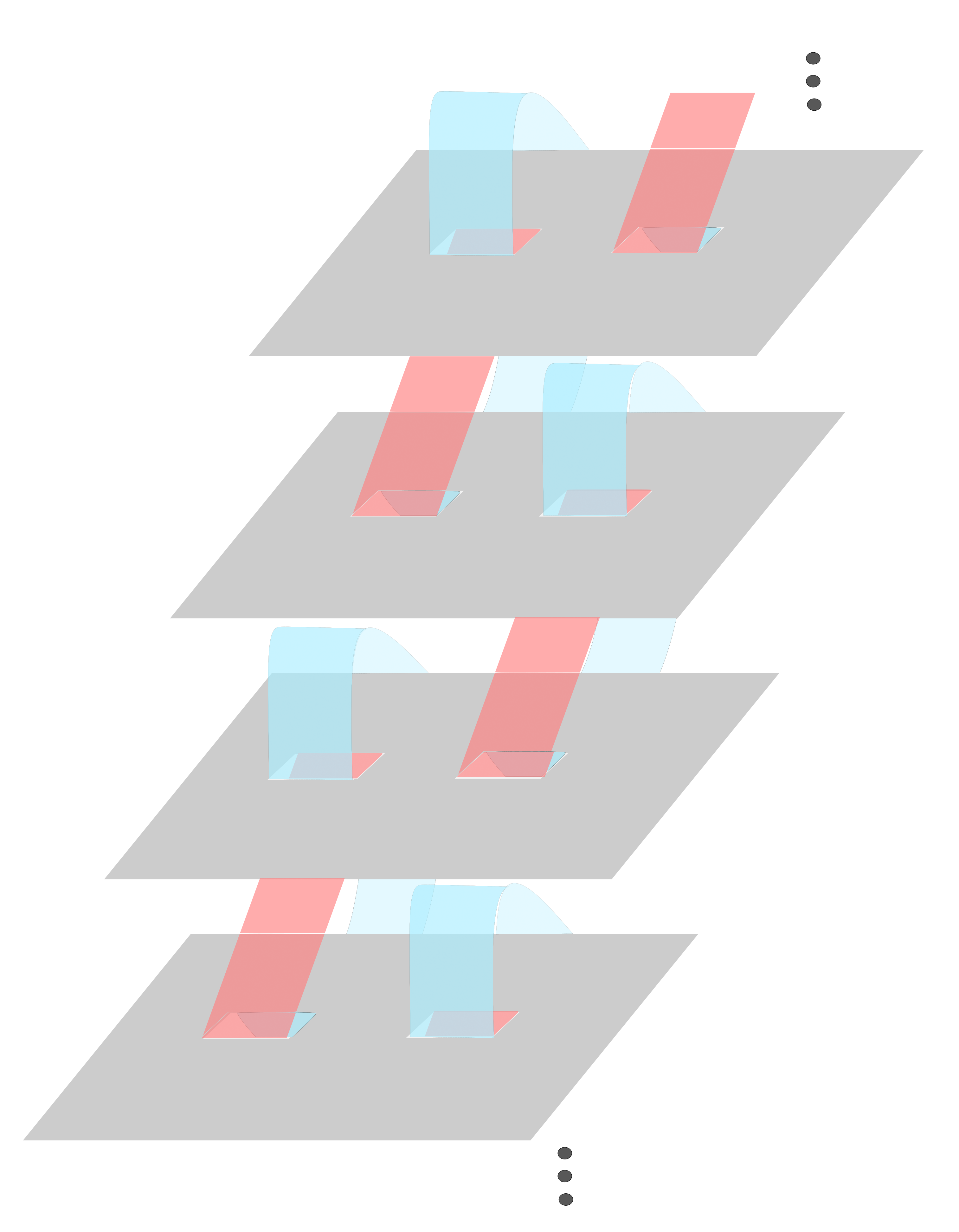}
\caption{\label{fig:ii} Riemann surface obtained for $\mathrm{Tr}\left(RR^{\dagger}\right)^n$ via the replica trick.}
\end{figure}

We recall that the mixed density matrix $\rho_{AB}$ for $A\cup B$ may be given in the field basis by the path integral on a single plane with cuts introduced along the intervals $A$ and $B$. The matrix product $RR^{\dagger}$ then amounts to taking two copies of $\rho_{AB}$ and pasting them together along the cuts on $B$.  The Riemann surface corresponding to $\mathrm{Tr}\left(RR^{\dagger}\right)^n$, is then constructed by taking the $n$ copies of $RR^{\dagger}$ and pasting the cuts along $A$ on the upper sheet of the $j^{th}$ copy with the cuts along $A$ on the lower sheet of the $(j+1)^{th}$ copy, where $j\in \{\mathbb{Z}\mod n\}$. The resulting surface is shown in Figure \ref{fig:ii}. It is not too difficult to see that this Riemann surface has a geometry of a genus-$1$ torus. 

It is well known that for CFTs the evaluation of the partition function on the Riemann surface is equivalent to computing the correlation function of the twist fields associated with this Riemann surface \cite{Cardy:2007mb, Castro-Alvaredo:2009yqb}. The idea behind the twist matrices is to consider the $n-$copies of the field on the plane and introduce twist fields such that the monodromy of the fields around these twist fields replicates the replica trick construction of the Riemann surface. We have the following relation for the partition function $Z_n$
\begin{equation} \label{eq3.1}
Z_n \propto \langle\mathcal{T}_{A,n}(u_1)\mathcal{T}_{A,n}(v_1)\mathcal{T}_{B,n}(u_2)\mathcal{T}_{B,n}(v_2)\rangle.
\end{equation}
Since the Riemann surface consists of $2n$ sheets, we consider $2n$ copies of the field on the single sheet. These fields satisfy the following monodromy conditions around the twist field,
\begin{equation} \label{eq3.2}
\begin{split}
\mathcal{T}_{A,n}\: :\: 2j\leftrightarrow 2j+1\: \mod n, \\
\mathcal{T}_{B,n}\: :\: 2j-1\leftrightarrow 2j\: \mod n,
\end{split}
\end{equation} 
where $j\in\{1,\cdots n\}$ and the field indices run over $\mathbb{Z} \mod 2n$. The scaling dimensions of these fields are $h_n=\frac{nc}{16}$, where $c$ is the central charge of the CFT.

Now, the case of the multiple disjoint intervals may be similarly considered, here $RR^{\dagger}$ is constructed by joining the two copies of $\rho_{AB}$ along $B=\cup_{i=1}^{N-1}B_i$, see Figure \ref{fig:i}. The Riemann surface corresponding to the $n^{th}$ moment of $RR^{\dagger}$ is then similarly constructed. As mentioned at the start of this section, in this case, however, the genus of the resulting Riemann surface is not fixed, but it is given by $g=n(N-1)-(n-1)$. In this work, we will be interested in the $n=2$ case only, and the genus here will be $g=2N-3$. The partition function is again given by
\begin{equation} \label{eq3.3}
Z_n\propto \langle\mathcal{T}_{A,n}(u_1)\mathcal{T}_{A,n}(v_1)\prod_{j=2}^{N}\mathcal{T}_{B,n}(u_j)\mathcal{T}_{B,n}(v_j)\rangle.
\end{equation}
The major goal of this work is to compute this correlation function for $n=2$ with an arbitrary number of intervals $N$ in the theory of compact boson. 
\section{Compact Boson} \label{section4}
Let us briefly introduce compact massless boson in $2d$. It is a conformally invariant theory with central charge $c=1$. Its action is given by,
\begin{equation} \label{eq4.1}
\mathcal{S}=\frac{1}{8\pi}\int\mathrm{d}^2x\,\partial_{\mu}\varphi\partial^{\mu}\varphi.
\end{equation}
The target space of the field $\varphi$ is compactified on a circle of radius $R$, that is, we identify $\varphi\sim \varphi+2\pi Rk$, where $k\in\mathbb{Z}$. Compact boson is the CFT of Luttinger liquids \cite{Giamarchi:2003ooa}, and the compactification radius $R$ is related to the Luttinger parameter $K$ via the relation $R=\sqrt{2/K}$.

Let us recall the results for the second R\'enyi entropy $(n=2)$ of multiple disjoint intervals, for two reasons: first, these results will be directly useful in later sections; and second, they provide a natural setting to introduce some of the techniques that we will employ throughout this work. While we do not derive these expressions here, we present the necessary definitions and outline key technical details in Appendix \ref{A}. The partition function $\mathcal{Z}_{N,2}$ for the second R\'enyi entropy is given by \cite{Coser:2013qda},
\begin{equation} \label{eq4.2}
\mathcal{Z}_{N,2}=c_{N,2}\left|\frac{\prod_{j>i=1}^{N}(u_j-u_i)(v_j-v_i)}{\prod_{j,i=1}^{N}(v_i-u_j)}\right|^{\frac{1}{4}}\mathcal{F}(x_1,\cdots,x_{2N-3}),
\end{equation}
where $c_{N,n}$ is a non-universal constant. The factor $\mathcal{F}$ is a function of the cross-ratios $x_i$, where 
\begin{equation} \label{eq4.4}
x_{2j-2}=\frac{\left(u_j-u_1\right)\left(v_N-u_N\right)}{\left(v_N-v_j\right)\left(u_N-u_1\right)},\qquad
x_{2j-1}=\frac{\left(v_j-u_1\right)\left(v_N-u_N\right)}{\left(v_N-v_j\right)\left(u_N-u_1\right)},
\end{equation}
with $j\in\{1,2,\cdots N\}$ and it is given by
\begin{equation} \label{eq4.5}
\mathcal{F}(x_1,\cdots,x_N)=\frac{\Theta\left(0|K\Pi\right)\Theta\left(0|\Pi/K\right)}{\left|\Theta\left(0|\Pi\right)\right|^2}.
\end{equation}
In the above equation, $\Theta$ is the Riemann Siegel theta function. A $k-$dimensional theta function is defined as
\begin{equation} \label{eq4.6}
\Theta\left[\begin{array}{l}
\boldsymbol{\varepsilon} \\
\boldsymbol{\delta}
\end{array}\right](\boldsymbol{\xi} \mid \Omega) \equiv \sum_{\boldsymbol{m} \in \mathbb{Z}^{k}} e^{i \pi(\boldsymbol{m}+\boldsymbol{\varepsilon})^t \cdot \Omega \cdot (\boldsymbol{m}+\boldsymbol{\varepsilon})+2 \pi i(\boldsymbol{m}+\boldsymbol{\varepsilon})^t \cdot(\boldsymbol{\xi}+\delta)},
\end{equation}
where the characteristics $\boldsymbol{\varepsilon}$, $\boldsymbol{\delta} \in \left(\mathbb{Z}/2\right)^k$ and $\boldsymbol{\xi} \in \mathbb{C}^k$. In eq.\eqref{eq4.6}, $\Omega$ is a $k\times k$ symmetric matrix with a positive definite imaginary part. For brevity the theta function in eq.\eqref{eq4.6} is denoted $\Theta(\boldsymbol{\xi}|\Omega)$ when $\varepsilon=\delta=0$. In eq.\eqref{eq4.5}, the $(N-1)\times(N-1)$ matrix  $\Pi$ is the Riemann period matrix of the Riemann surface $\tilde{\Sigma}_{N}$ parametrised by the algebraic curve
\begin{equation} \label{eq4.7}
u^2=\prod_{j=0}^{2N-2}(z-x_j),
\end{equation}
see appendix \ref{A} for more details. This is just the Riemann surface associated with the twist field correlation function $\langle\prod_{j=0}^{2N-1}\mathcal{T}_2(x_j)\rangle$, which we get after a global conformal transformation
\begin{equation*}
w=\frac{(z-u_1)(v_N-u_N)}{(v_N-z)(u_N-u_1)}.
\end{equation*} 
The partition function $\mathcal{Z}_{N,2}$ may also be written as $\mathcal{Z}_{N,2}^{qu}\mathcal{Z}_{N,2}^{cl}$, where $qu$ and $cl$ stand for quantum and classical part of the twist field correlation functions discussed in Section \ref{section3}, these are given by
\begin{align} 
\mathcal{Z}_{N,2}^{qu}&\propto \left|\frac{\prod_{j>i=1}^{N}(u_j-u_i)(v_j-v_i)}{\prod_{j,i=1}^{N}(v_i-u_j)}\right|^{\frac{1}{4}}\frac{1}{\sqrt{\det(\text{Img}\Pi)}\left|\Theta\left(0|\Pi\right)\right|^2},\label{eq4.8}\\
\mathcal{Z}_{N,2}^{cl}&\propto \sqrt{\det(\text{Img}\Pi)}\Theta\left(0|K\Pi\right)\Theta\left(0|\Pi/K\right). \label{eq4.9}
\end{align}
The proportionality constant in $\mathcal{Z}_{N,2}$ is fixed by the small intervals and large distance limit, i.e. $\frac{\ell_i}{d_j}<<1$, for all $i,j$. In this limit, the reduced density matrix becomes separable, and it is just the product of the reduced density matrices for $N$ single intervals. This constant should not be confused with the non-universal constant $c_{N,2}$, which depends on the lattice theory. In the following subsections, we will determine the quantum and classical part of the twist field correlation function for $2$-R\'enyi CCNR negativity.

\subsection{Quantum Part} \label{section4.1}
As discussed in Section \ref{section3}, we need to evaluate the twist field correlation function, eq.\eqref{eq3.3} to find the R\'enyi CCNR negativity. We also argued that in the case $N>2$ the associated Riemann surface doesn't have a fixed genus as $n$ varies, as opposed to the $N=2$ case. This makes the evaluation of the twist field correlation function complicated. The situation is even more complicated for $n>2$ since the twist fields $\mathcal{T}_{A,n}$ and $\mathcal{T}_{B,n}$ become non-abelian and hence cannot be simultaneously diagonalised. For this reason, in the present work, we will only focus on the $n=2$ case and hope to come back to the more general case in a later work.

In the $n=2$ case, we may diagonalise the twist field $\mathcal{T}_{A}$ and $\mathcal{T}_{B}$ (we drop the $n=2$ in the subscript notation for brevity) simultaneously,
\begin{equation} \label{eq4.1.1}
\mathcal{T}_{A}=
\left[
\begin{array}{cccc}
0 & 0 & 0 & 1\\
0 & 0 & 1 & 0\\
0 & 1 & 0 & 0\\
1 & 0 & 0 & 0
\end{array}
\right],
\qquad
\mathcal{T}_{B}=
\left[
\begin{array}{cccc}
0 & 1 & 0 & 0\\
1 & 0 & 0 & 0\\
0 & 0 & 0 & 1\\
0 & 0 & 1 & 0
\end{array}
\right]
\end{equation}
by the changing the field basis from $\varphi$ to $\tilde{\varphi}$,
\begin{equation} \label{eq4.1.2}
\left[\tilde{\varphi}\right]=
\frac{1}{2}\left[
\begin{array}{rrrr}
-1 & 1 & -1 & 1\\
1 & 1 & 1 & 1\\
-1 & -1 & 1 & 1\\
1 & -1 & -1 & 1
\end{array}
\right]
\left[\varphi\right].
\end{equation}
The diagonalised twist fields are then given by,
\begin{equation} \label{eq4.1.3}
\mathcal{T}_{A}=
\left[
\begin{array}{cccc}
-1 & 0 & 0 & 0\\
0 & 1 & 0 & 0\\
0 & 0 & -1 & 0\\
0 & 0 & 0 & 1
\end{array}
\right],
\qquad
\mathcal{T}_{B}=
\left[
\begin{array}{cccc}
-1 & 0 & 0 & 0\\
0 & 1 & 0 & 0\\
0 & 0 & 1 & 0\\
0 & 0 & 0 & -1
\end{array}
\right].
\end{equation}
In this basis, we'll then need to evaluate the following correlation function,
\begin{equation} \label{eq4.1.4}
\begin{split}
Z_{N} \propto \langle\mathcal{T}_{A,1/2}(u_1)\mathcal{T}_{A,1/2}(v_1)\prod_{j=2}^{N}\mathcal{T}_{B,1/2}(u_j)\mathcal{T}_{B,1/2}(v_j)\rangle\langle\prod_{j=2}^{N}&\mathcal{T}_{B,1/2}(u_j)\mathcal{T}_{B,1/2}(v_j)\rangle\\
&\langle\mathcal{T}_{A,1/2}(u_1)\mathcal{T}_{A,1/2}(v_1)\rangle.
\end{split}
\end{equation}
Here we have introduced the diagonalised twist fields $\mathcal{T}_{J,k/n}^i$, with $J\in\{A,B\}$, around which the fields $\tilde{\varphi}^j$ satisfies the following monodromy relations,
\begin{equation}
\mathcal{T}_{J,k/n}^i: \tilde{\varphi}^j \to e^{i2\pi\frac{k}{n}}\delta_{ij}\tilde{\varphi}^j +(1-\delta_{ij})\tilde{\varphi}^j.
\end{equation}
The superscript index $i$ is omitted in eq.\eqref{eq4.1.4}, as it no longer plays a role in the subsequent calculations. However, such a change of basis will result in non-trivial winding for the field $\tilde{\varphi}$ around the twist fields. This will make the calculation of the classical part complicated, and for this reason, we will calculate the classical part in subsection \ref{section4.3} directly from eq.\eqref{eq3.3}. The quantum part, however, is independent of the winding and depends upon only the global monodromy of the fields around the twist fields \cite{Dixon:1986qv}, hence it becomes easier to compute it this way.

\begin{figure}
\centering % \begin{center}/\end{center} takes some additional vertical space
%\includegraphics[width=.5\textwidth]{tikz1.pdf}
% "\includegraphics" is very powerful; the graphicx package is already loaded
\includegraphics[width=1\textwidth]{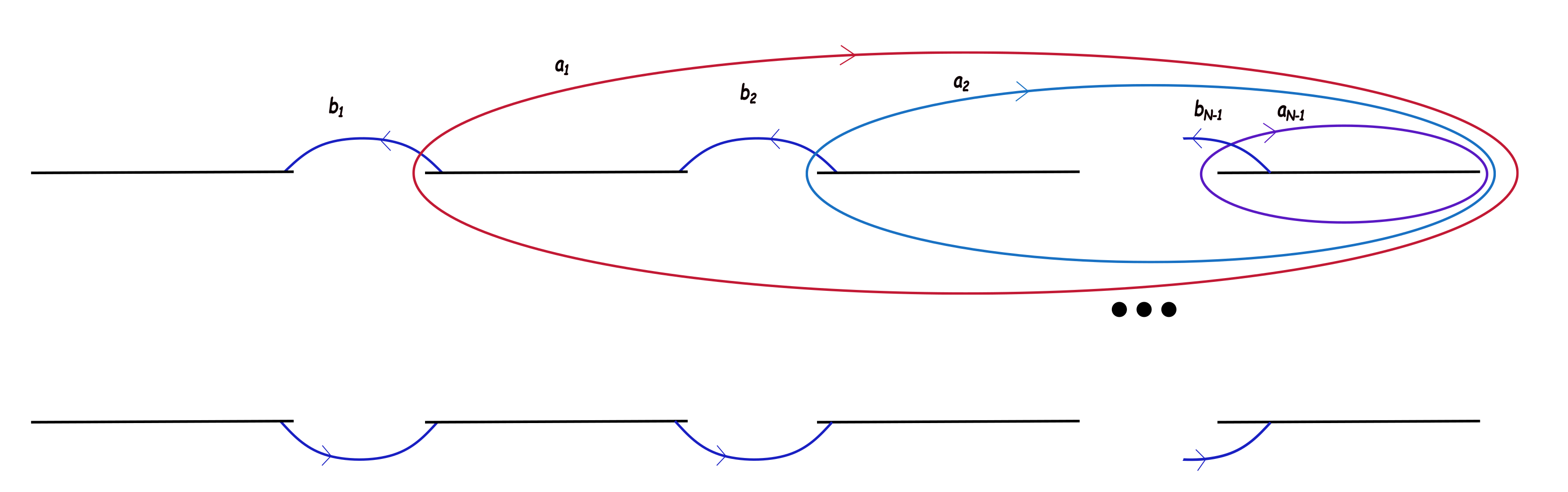}
\caption{\label{fig:4i} Homology basis for general N: Homology basis for $\Sigma_N$ is shown above, the horizontal lines represent branch cuts.}
\end{figure}

For the reasons that will become clear later, we use a global conformal transformation $w$ to fix the points $u_1\to-\infty$, $v_1\to 0$, and $v_N\to 1$,
\begin{equation} \label{eq4.1.5}
w=\frac{(z-v_1)(v_N-u_1)}{(z-u_1)(v_N-v_1)},
\end{equation}
and under $w$ we have the following map for the branch points,
\begin{equation} \label{eq4.1.6}
y_{2j-3}=\frac{\left(u_j-v_1\right)\left(v_N-u_1\right)}{\left(u_j-u_1\right)\left(v_N-v_1\right)}\qquad
y_{2j-2}=\frac{\left(v_j-v_1\right)\left(v_N-u_1\right)}{\left(v_j-u_1\right)\left(v_N-v_1\right)},
\end{equation}
where $j\in\{1,\ldots,N\}$. Under this map, we have the following expression for the twist field correlation function
\begin{equation} \label{eq4.1.7}
\begin{split}
\langle\mathcal{T}_{A,1/2}(u_1)\cdots\mathcal{T}_{B,1/2}(v_N)\rangle=&\left(\frac{(v_N-u_1)^{N-1}(v_1-u_1)^{N-2}}{(v_N-v_1)^{N-1}\prod_{j=2}^N(u_j-u_1)(v_j-v_1)}\right)^{\frac{1}{4}}\\
&\hspace{1.2in}\langle\mathcal{T}_{A,1/2}(-\infty)\cdots\mathcal{T}_{B,1/2}(1)\rangle,
\end{split}
\end{equation}
where we have introduced $\mathcal{T}_{A,1/2}(-\infty)=\lim_{w\to-\infty}w^{1/4}\mathcal{T}_{A,1/2}(w)$. The quantum part of the twist field correlation function is known in literature \cite{Dixon:1986qv, Verlinde:1986kw, Dijkgraaf:1987vp,Alvarez-Gaume:1986rcs}, see specifically \cite{Zamolodchikov:1987ae} for the present calculations. To compute the quantum part, first, let us consider the Riemann surface corresponding to the twist fields on the right side of the above equation. This surface is parametrised by the curve
\begin{equation} \label{eq4.1.8}
y^2=\prod_{i=0}^{2N-2}(w-y_i).
\end{equation}
As discussed in Appendix \ref{A} to find the period matrix, we introduce a homology basis as shown in Figure \ref{fig:4i}. We choose the basis of holomorphic differentials of the first kind $\mathrm{d}\mu_j$ to be
\begin{equation} \label{eq4.1.9}
\mathrm{d}\mu_j(w)=\frac{w^{j-1}}{y}\mathrm{d}w.
\end{equation} 
The Riemann period matrix $\Omega_{N}$, is then determined by using eq.\eqref{eq4.1.9} in eq.\eqref{eqA.6}-\eqref{eqA.7}. We checked that the period matrix $\Omega_{N}$ obtained here is the same as one would obtain in eq.\eqref{eq4.5} using the conventions of appendix \ref{A}.
The quantum part of the twist field correlation function is given by
\begin{equation} \label{eq4.1.10}
\langle\mathcal{T}_{1/2}(-\infty)\cdots\mathcal{T}_{1/2}(1)\rangle_{\text{qu}}\propto\left|\frac{\prod_{k>l}^{N}(y_{i_k}-y_{i_l})(y_{j_k}-y_{j_l})}{\prod_{k,l}^{N}(y_{i_k}-y_{j_k})}\right|^{\frac{1}{4}}\frac{1}{\sqrt{\det(\text{Img}\Omega_N)}\Theta[\boldsymbol{e}](0|\Omega_N)^2},
\end{equation} 
where the branch points groups into two disjoint sets $\{y_{i_k}\}$ and $\{y_{j_k}\}$, with each set having $N$ points \cite{mumford2007tata}. Note that we have dropped the subscripts $A$ and $B$ above since the quantum part is independent of $A$ and $B$. This grouping depends upon the choice of the characteristics $\boldsymbol{e}$. We set $\boldsymbol{e}=0$, then one of the sets, say $\{y_{i_k}\}$ is just the collections of branch points that are the zeros of theta function $\Theta(\boldsymbol{\alpha}|\Omega_N)$ under the Abel-Jacobi map $\boldsymbol{\alpha}$ along with the branch point $0$. The Abel-Jacobi map is defined from the Riemann surface $\Sigma_N$ to its Jacobian torus $J(\Sigma_N)$ and is a $(N-1)$ dimensional vector in the present case given by
\begin{equation} \label{eq4.1.11}
\alpha_j(w)=\int_0^{w}\mathrm{d}\nu_j(w').
\end{equation}
The normalised holomorphic differentials $\mathrm{d}\nu_j$ are obtained by using eq.\eqref{eq4.1.9} in eq.\eqref{eqA.6}, using the homology basis given by Figure \ref{fig:4i}. To determine this grouping, we first note Theta function relation
\begin{equation} \label{eq4.1.12}
\Theta\left[\begin{array}{l}
\boldsymbol{0} \\
\boldsymbol{0}
\end{array}\right](\Omega_N\cdot g_1+g_2 \mid \Omega_N)=e^{-i\pi g_1^t\cdot\Omega_N\cdot g_1-i2\pi g_1^t\cdot g_2}\Theta\left[\begin{array}{l}
\boldsymbol{g1} \\
\boldsymbol{g2}
\end{array}\right](0\mid \Omega).
\end{equation}
The theta function vanishes for all odd characteristics, that is, when the characteristics $g_1$ and $g_2$ satisfy $4g_1^t\cdot g_2=1\mod 2$. We may thus conclude that $\boldsymbol{\alpha}(y_j)=\Omega_N\cdot g_1+g_2$ is a zero of theta function if this condition is satisfied. We have
\begin{equation} \label{eq4.1.13}
\begin{split}
(g_1)_i=\frac{1}{2}u(j-i)\text{ for } y_{2j-1},y_{2j},\\
(g_2)_i=\frac{1}{2}\left(\delta_{1,i}-\delta_{j,i}\right)\text{ for } y_{2j-1},y_{2j-2},
\end{split}
\end{equation}
where $u(x)$ is unit step function and vanishes only for $x<0$. Therefore, we find that the $y_{2j}$ are the zeros of the theta function, and the two sets are given by
\begin{equation} \label{eq4.1.14}
\{y_{i_k}\}=\{0,y_2,\cdots y_{2N-2}\}, \qquad \{y_{j_k}\}=\{-\infty,y_1,\cdots y_{2N-3}\}.
\end{equation}
Using this in eq.\eqref{eq4.1.10} and subsequently substituting eq.\eqref{eq4.1.10} in eq.\eqref{eq4.1.7} we obtain the following expression for the quantum part,
\begin{equation} \label{eq4.1.15}
\langle\mathcal{T}_{1/2}(u_1)\cdots\mathcal{T}_{1/2}(v_N)\rangle_{\text{qu}}\propto\left|\frac{\prod_{j>i=1}^{N}(u_j-u_i)(v_j-v_i)}{\prod_{j,i=1}^{N}(v_i-u_j)}\right|^{\frac{1}{4}}\frac{1}{\sqrt{\det(\text{Img}\Omega_N)}\Theta(0|\Omega_N)^2},
\end{equation}
where we absorbed the algebraic factor of eq.\eqref{eq4.1.7} in the above expression. We mention that this is the same as the corresponding quantum part in eq.\eqref{eq4.8}, as one would expect.

\begin{figure}
\centering % \begin{center}/\end{center} takes some additional vertical space
%\includegraphics[width=.5\textwidth]{tikz1.pdf}
% "\includegraphics" is very powerful; the graphicx package is already loaded
\includegraphics[width=1\textwidth]{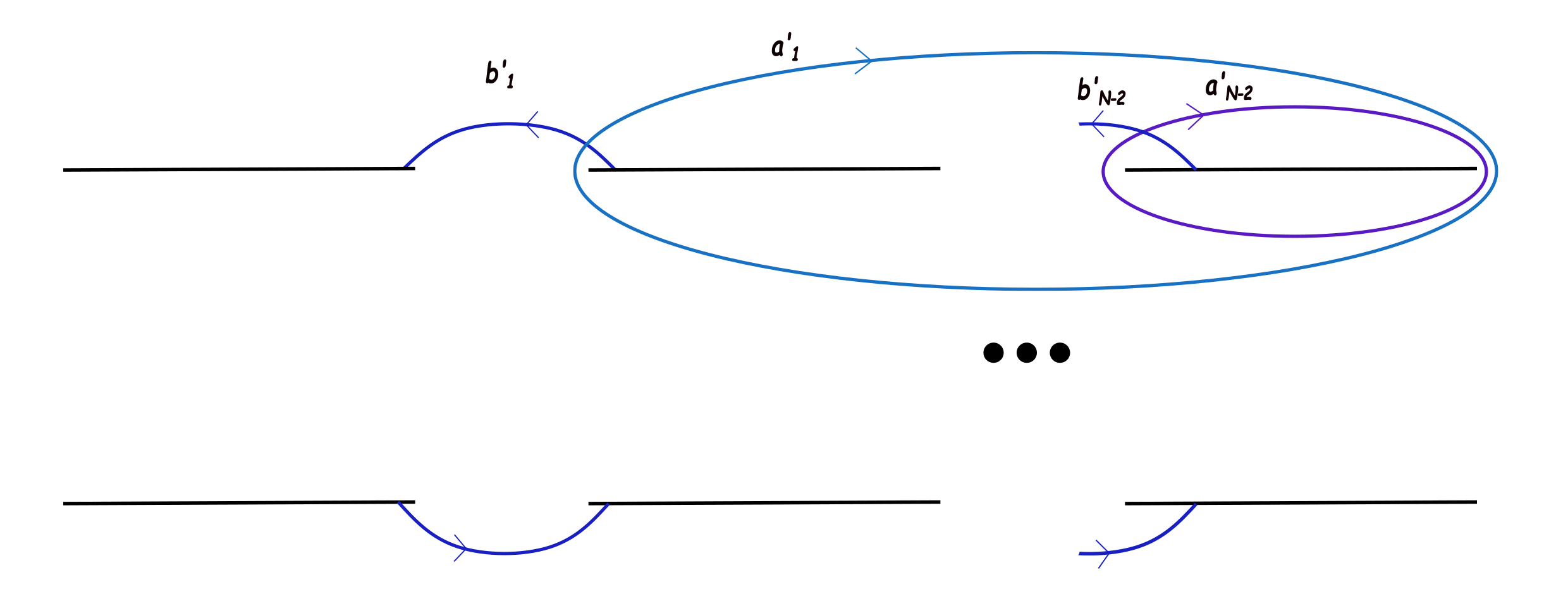}
\caption{\label{fig:4ij} Homology basis for $\Sigma'_{N-1}$: Homology basis for $\Sigma'_{N-1}$ is shown above, the horizontal lines represent branch cuts. The loops may be recognised as $a'_{j}=a_{j+1}$ and $b'_{j}=b_{j+1}$, where $a$, and $b$ are the homology basis on $\Sigma_N$.}
\end{figure}

The quantum part of the twist field correlation function $\langle\prod_{j=2}^{N}\mathcal{T}_{B,1/2}(u_j)\mathcal{T}_{B,1/2}(v_j)\rangle$ is similarly evaluated. In this case the relevant Riemann surface $\Sigma'_{N-1}$ is parametrised by the curve
\begin{equation} \label{eq4.1.16}
y'^2=\prod_{j=1}^{2N-2}(w-y_j).
\end{equation}
We choose a holomorphic basis similar to the one shown in Figure \ref{fig:4ij}. The Riemann surface $\Sigma'_{N-1}$ is just the surface $\Sigma_{N}$ with the branch cut $(-\infty,0)$ removed. The basis of the holomorphic differentials of the first kind may be chosen as
\begin{equation}
\mathrm{d}\mu'_{j}=\frac{w^{j-1}}{y'}\mathrm{d}w.
\end{equation}
The rest of the calculations follow similarly, and we get the following quantum part for the twist field correlation function
\begin{equation} \label{eq4.1.17}
\langle\prod_{j=2}^{N}\mathcal{T}_{1/2}(u_j)\mathcal{T}_{1/2}(v_j)\rangle_{\text{qu}}\propto\left|\frac{\prod_{j>i=2}^{N}(u_j-u_i)(v_j-v_i)}{\prod_{j,i=2}^{N}(v_i-u_j)}\right|^{\frac{1}{4}}\frac{1}{\sqrt{\det(\text{Img}\tau_{N-1})}\Theta(0|\tau_{N-1})^2},
\end{equation}
where $\tau_{N-1}$ is the period matrix of the Riemann surface $\Sigma'_{N-1}$ and it is the same as one would get in eq.\eqref{eq4.5} for the same set of twist fields. The remaining part of eq.\eqref{eq4.1.4} is simple to evaluate and is given by
\begin{equation} \label{eq4.1.18}
\langle\mathcal{T}_{1/2}(u_1)\mathcal{T}_{1/2}(v_1)\rangle=\frac{1}{(v_1-u_1)^{\frac{1}{4}}}.
\end{equation}
We finally obtain the quantum part of eq.\eqref{eq4.1.4} from eq.\eqref{eq4.1.15}, eq.\eqref{eq4.1.17}, and eq.\eqref{eq4.1.18} to be
\begin{equation}
\begin{split}
Z_{N}^{\text{qu}}\propto \left|\frac{\prod_{j>i=2}^{N}(u_j-u_i)(v_j-v_i)}{(v_1-u_1)\prod_{j,i=2}^{N}(v_i-u_j)}\right|^{\frac{1}{2}}&   \left|\frac{\prod_{j=2}^{N}(u_j-u_1)(v_j-v_1)}{\prod_{j=2}^{N}(v_j-u_1)(u_j-v_1)}\right|^{\frac{1}{4}}\times\\
&\frac{1}{\sqrt{\det(\text{Img }\tau_{N-1})}\Theta(0|\tau_{N-1})^2}\frac{1}{\sqrt{\det(\text{Img }\Omega_{N})}\Theta(0|\Omega_{N})^2},
\end{split}
\end{equation}

\subsection{Classical Part} \label{section4.2}
\begin{figure}
\centering % \begin{center}/\end{center} takes some additional vertical space
%\includegraphics[width=.5\textwidth]{tikz1.pdf}
% "\includegraphics" is very powerful; the graphicx package is already loaded
\includegraphics[width=1\textwidth]{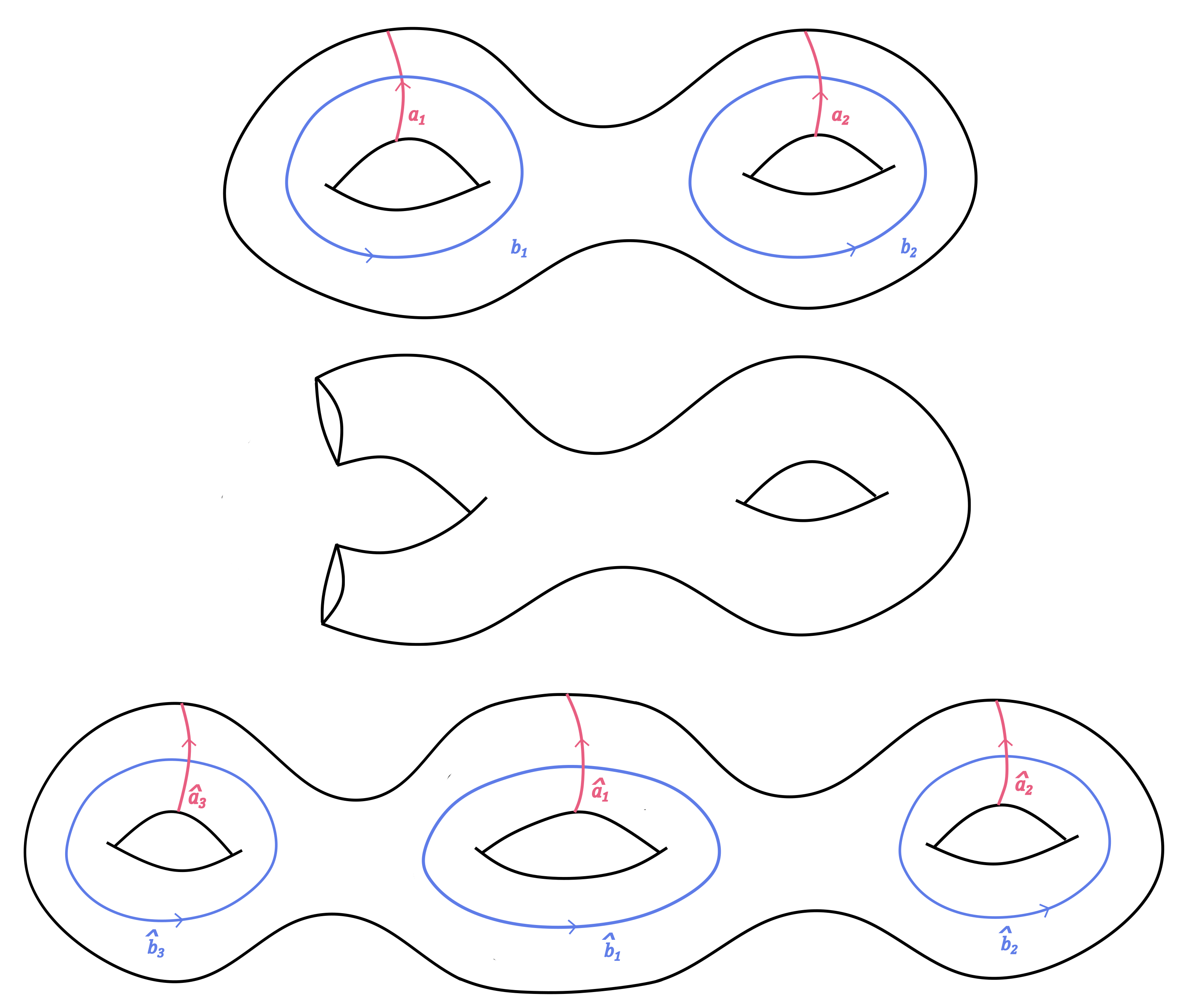}
\caption{\label{fig:4ii} Riemann surface $\hat{\Sigma}_N$: $\hat{\Sigma}_N$ (bottom) is a double cover of ${\Sigma}_N$ (top), and is constructed by taking two copies of $\Sigma_N$, cutting then along the $a_1$ loop and pasting them together along this cut.}
\end{figure}

We will calculate the classical part using a more direct approach, that is by computing the period matrix of the Riemann surface resulting from the replica trick or the twist field correlation function of eq.\eqref{eq3.3}. The key observation is that the Riemann surface for $\mathrm{Tr}(RR^{\dagger})^2$ is obtained by taking two copies of $\Sigma_N$, cutting both of them along the loop $a_1$, see Figure \ref{fig:4i} for the homology basis, and pasting the two copies along this cut, see Figure \ref{fig:4ii} for illustration.

We mention that this construction is similar to the one which occurs in the evaluation of partition functions for $Z_2$-orbifold compact boson \cite{Dijkgraaf:1987vp, Miki:1987mp}. While computing the contribution of the twisted sectors to the partition function, one introduces a double covering of the torus on which the partition function is computed. Our construction above is the double covering of $\Sigma_N$, for the case when the twisted boundary condition is imposed along the loop $b_1$. We will use this formalism of double covering to compute the period matrix of the desired Riemann surface.

Let us denote by $\hat{\Sigma}_N$ the desired double cover of $\Sigma_N$. Let's briefly discuss some properties of $\hat{\Sigma}_N$. First from the Riemann-Hurwitz formula, $\hat{\Sigma}_N$ has a genus $2(N-1)-1=2N-3$, as already argued in Section \ref{section3}. The surface $\hat{\Sigma}_N$ naturally endows a homology basis from $\Sigma_N$, as shown in Figure \ref{fig:4ii}. To describe the construction of this homology basis, we note that there is a conformal automorphism $\alpha$ on $\hat{\Sigma}_N$, satisfying $\alpha\circ \alpha=e$, where $e$ is the identity map. Under this conformal automorphism, the homology basis satisfies the following relations,
\begin{align}
\alpha\circ \hat{b}_1=\hat{b}_1, &\qquad \alpha\circ \hat{a}_1=\hat{a}_1, \label{eq4.2.1}\\
\alpha\circ \hat{b}_j=\hat{b}_{2N-3-j}, &\qquad \alpha\circ \hat{a}_j=\hat{a}_{2N-3-j}, \label{eq4.2.2}
\end{align}
where $j\in\{2,\ldots, 2N-3\}$. The basis cycles $\hat{a}_i$ and $\hat{b}_i$, where $i\in\{1,2\cdots N-1\}$, are chosen such that under the projection $\hat{\Sigma}_N/\alpha=\Sigma_N$, they project onto $a_i$ and $b_i$, respectively. In particular, the loop $\hat{b}_1$ is formed by joining two $b_1$ loops, as is also illustrated in Figure \ref{fig:4ii}. 

The next step is to find a basis of holomorphic differential of the first kind $\mathrm{d}\hat{w}_j$ on $\hat{\Sigma}_N$. The lift of normalised holomorphic differentials $\mathrm{d}\nu_j$ on $\Sigma_N$ gives the first $N-1$ holomorphic differentials of the first kind on $\hat{\Sigma}_N$. The lift of differentials $\mathrm{d}\nu_j$ are no longer normalised on $\hat{\Sigma}_N$ since they satisfy
\begin{align}
&\oint_{\hat{a}_1}\mathrm{d}\hat{w}_j=\delta_{1,j},\label{eq4.2.3}\\
&\oint_{\hat{a}_i}\mathrm{d}\hat{w}=\oint_{\hat{a}_{2N-3-i}}\mathrm{d}\hat{w}=\delta_{i,j}, \label{eq4.2.4}
\end{align}
where $i,j\in\{1,2\cdots N-1\}$. The remaining $N-2$ holomorphic differentials of the first kind are given by Prym differentials on $\hat{\Sigma}_N$. Prym differentials $\mathrm{d}\hat{w}_j$ (with $j\in\{N,\ldots, 2N-3\}$) by construction are odd under the conformal automorphism $\alpha$ and span a $N-2$ dimensional space. In terms of $\Sigma_N$ they have a vanishing period around the loop ${a}_1$, they are double valued around the loop $b_1$ (i.e. anti-periodic) and single valued around the remaining loop basis (see \cite{Dijkgraaf:1987vp} for more details). The differentials $\mathrm{d}\hat{w}_j$ satisfy the following relations under the conformal automorphism,
\begin{align}
\alpha^*\circ\mathrm{d}\hat{w}_j&=\mathrm{d}\hat{w}_j, \qquad j\in\{1,\ldots, N-1\}\label{eq4.2.5}\\
\alpha^*\circ\mathrm{d}\hat{w}_j&=-\mathrm{d}\hat{w}_j, \qquad j\in\{N,\ldots, 2N-3\}. \label{eq4.2.6}
\end{align}
In the remaining part of this subsection, we will give the construction of $\hat{\Sigma}_N$ from $\Sigma_N$ following the references \cite{rauch1970theta, farkas1970period}, and utilise the above formalism to compute the classical part of the twist field correlation function.

\begin{figure}
\centering % \begin{center}/\end{center} takes some additional vertical space
%\includegraphics[width=.5\textwidth]{tikz1.pdf}
% "\includegraphics" is very powerful; the graphicx package is already loaded
\includegraphics[width=1\textwidth]{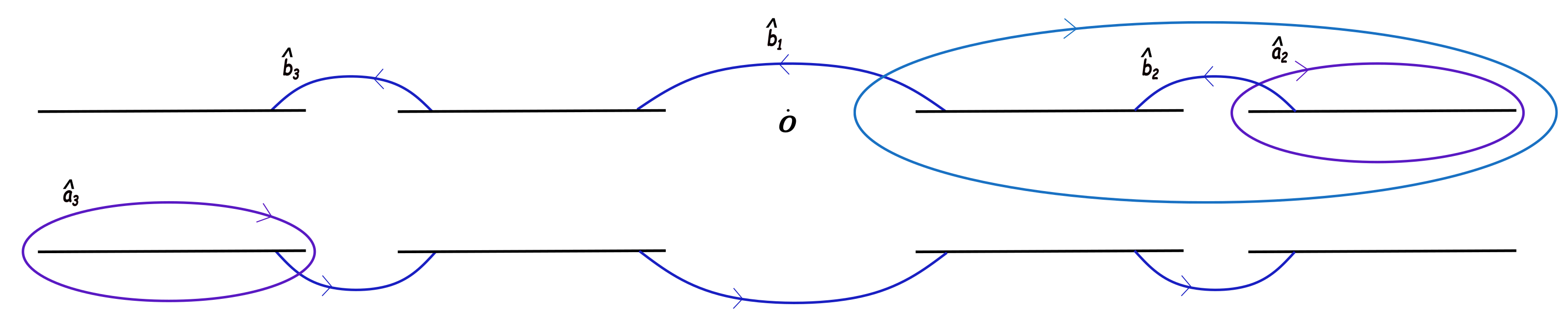}
\caption{\label{fig:4iii} Homology basis on $\hat{\Sigma}_N$: The homology basis $\hat{a}$, and $\hat{b}$ for $N=3$ is shown in the figure.}
\end{figure}

To construct the Riemann surface $\hat{\Sigma}$, we make the substitution $\zeta^2=w$ in eq.\eqref{eq4.1.16}, and obtain the hyperelliptic curve,
\begin{equation} \label{eq4.2.7}
\hat{y}^2=\prod_{i=1}^{2N-2}\left(\zeta-\sqrt{y_i}\right)^{\frac{1}{2}}\left(\zeta+\sqrt{y_i}\right)^{\frac{1}{2}}.
\end{equation}
The Riemann surface obtained from the branched $\zeta$ planes is a model for the surface $\hat{\Sigma}_N$, see \cite{rauch1970theta, farkas1970period}. The conformal automorphism $\alpha$ corresponds to a reflection through $\zeta=0$, followed by an interchange of two $\zeta$ sheets, i.e $(\hat{y},\zeta)\to(-\hat{y},-\zeta)$. Note that due to the sheet interchange $\zeta=0$ is not a fixed point, more generally, there is no fixed point under this conformal automorphism and hence $\hat{\Sigma}_N/\alpha$ is a manifold and not an orbifold. A homology basis for $\hat{\Sigma}_N$ satisfying eq.\eqref{eq4.2.1}-\eqref{eq4.2.2} is shown in Figure \ref{fig:4iii} for $N=3$ and can be generalised to arbitrary values of $N$ similarly. 

We delegate the computation of the Abelian differentials of the first kind and other necessary quantities to determine the period matrix to Appendix \ref{B}. The period matrix $T$ on $\hat{\Sigma}_N$ is obtained by using eq.\eqref{eqB.6}-\eqref{eqB.7} in eq.\eqref{eqA.6}-\eqref{eqA.7},
\begin{equation} \label{eq4.2.8}
{T}_{i j}=\left\{\begin{array}{rl}
2\Omega_{1,1} & \hspace{0.3in}  i=j=1, \\
\Omega_{i,1} & \hspace{0.3in} i \in\{2, \ldots, N-1\}, \\
\Omega_{2N-1-i,1} & \hspace{0.3in} i \in\{N, \ldots, 2N-3\}, \\
\Omega_{1,j} & \hspace{0.3in} j \in\{2, \ldots, N-1\}, \\
\Omega_{1,2N-1-j} & \hspace{0.3in} j \in\{N, \ldots, 2N-3\}, \\
\frac{1}{2}\left(\Omega_{i,j}+\tau_{i-1,j-1}\right) & \hspace{0.3in} i,j \in\{2, \ldots, N-1\}, \\
\frac{1}{2}\left(\Omega_{i,2N-1-j}-\tau_{i-1,2N-2-j}\right) & \hspace{0.3in} i \in\{2, \ldots, N-1\},\text{ }j\in\{N,\ldots, 2N-3\} ,\\
\frac{1}{2}\left(\Omega_{2n-1-i,j}-\tau_{2N-2-i,j-1}\right) & \hspace{0.3in} i \in\{N,\ldots, 2N-3\},\text{ }j\in\{2, \ldots, N-1\}, \\
\frac{1}{2}\left(\Omega_{2n-1-i,2n-1-j}+\tau_{2N-2-i,2N-2-j}\right) & \hspace{0.3in} i,j \in\{N,\ldots, 2N-3\}, 
\end{array}\right.
\end{equation}
where $\Omega$ and $\tau$ are the period matrices of $\Sigma_N$ and $\Sigma'_{N-1}$ respectively. We have dropped subscript denoting the dependence on $N$ in the notation above for brevity as well. As an example the period matrix $T$ for $N=3$ is given by
\begin{equation} \label{eq4.2.9}
T=
\left[
\begin{array}{ccc}
2\Omega_{1,1} & \Omega_{1,2} & \Omega_{1,2}\\
\Omega_{2,1} & \frac{1}{2}\left(\Omega_{2,2}+\tau\right) & \frac{1}{2}\left(\Omega_{2,2}-\tau\right)\\
\Omega_{2,1} & \frac{1}{2}\left(\Omega_{2,2}-\tau\right)& \frac{1}{2}\left(\Omega_{2,2}+\tau\right)\\
\end{array}
\right],
\end{equation}
We will use this expression in the next subsection to compare our results with some numerics. We mention that since $\Omega$ and $\tau$ in eq.\eqref{eq4.2.8} are completely imaginary, $T$ is also completely imaginary. Finally, we also need to compute the determinant of $T$ to determine the classical part. We find,
\begin{equation} \label{eq4.2.10}
\det(T)=2\det(\Omega)\det(\tau),
\end{equation}
to see this, we will use the row-wise linearity property of the determinants. We split each row of $T$, except for the first one, into two parts, one containing only $\Omega$ terms and the other only $\tau$. We get $(2N-2)^2$ determinants, however, most of these vanish since the rows are no longer linearly independent. Consider the determinant where the first $N-1$ rows depend on $\Omega$, and the remaining have $\tau$ dependence, it is not that difficult to see that this determinant is $2^{-N+3}\det(\Omega)\det(\tau)$. All the other non-vanishing determinants just correspond to some row interchange between the $\Omega$ dependent row and $\tau$ dependent row, up to a negative sign for the $\tau$ dependent row. All such contributions are again $2^{-N+3}\det(\Omega)\det(\tau)$, and there are in total $2^{N-2}$ of these contributions, and hence we have the result of eq.\eqref{eq4.2.9}.

Finally, the classical part of the twist field correlation function of eq.\eqref{eq3.3} is given by
\begin{equation} \label{eq4.2.11}
Z_{N}^{cl}\propto \sqrt{\mathrm{det}(\mathrm{Img\Omega_N})\mathrm{det}(\mathrm{Img\tau_{N-1}})}\Theta\left(0|KT\right)\Theta\left(0|T/K\right).
\end{equation} 

\subsection{$2$-R\'enyi CCNR Negativity} \label{section4.3}
\begin{figure}
\centering % \begin{center}/\end{center} takes some additional vertical space
%\includegraphics[width=.5\textwidth]{tikz1.pdf}
% "\includegraphics" is very powerful; the graphicx package is already loaded
\includegraphics[width=.8\textwidth]{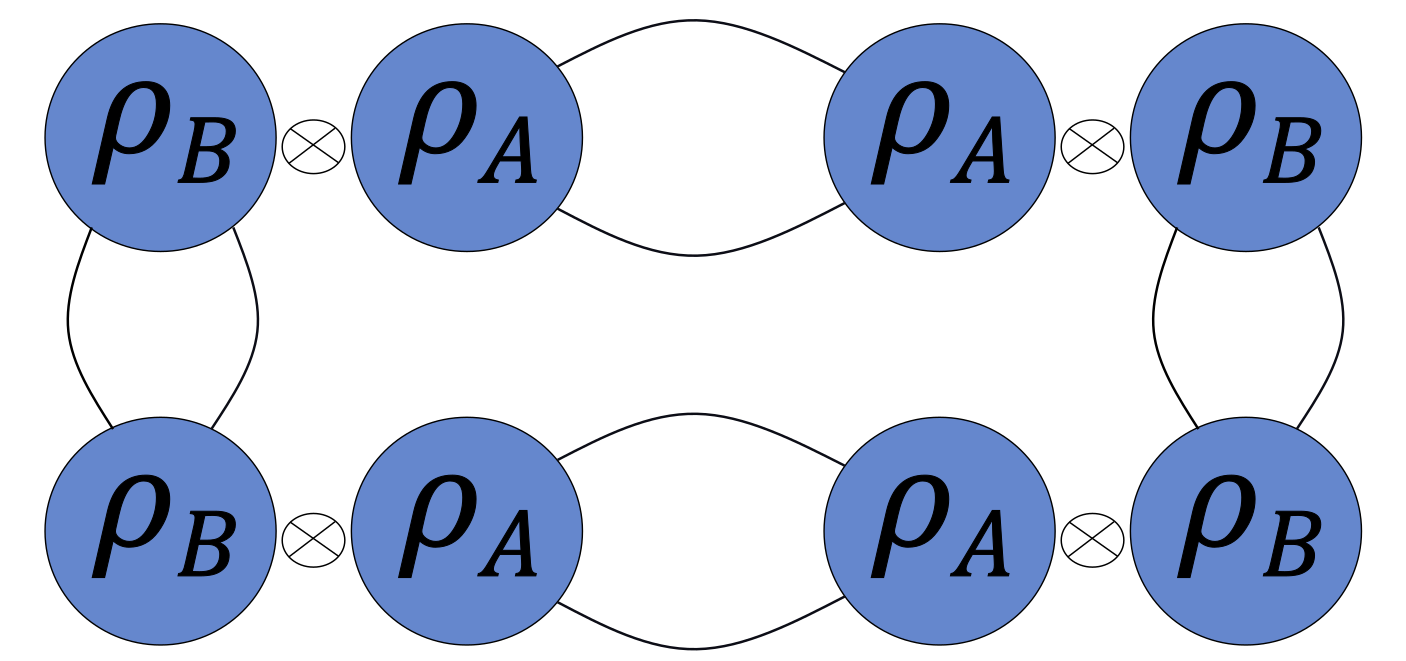}
\caption{\label{fig:4iv} Tensor Diagram for Far distance limit: The density matrix becomes separable in the far distance limit $\rho=\rho_A\otimes\rho_B$, the tensor diagram is shown for $\mathrm{Tr}(RR^\dagger)^2$}.
\end{figure}

To fix the constant of proportionality in $Z_N=Z_N^{cl}Z_N^{qu}$, we consider the limit of faraway intervals, i.e. $\frac{\ell_i}{d_j}<<1$ for all $i,j\in\{1,\ldots,N\}$. In this limit, the density matrix $\rho$ will become separable in all intervals, and therefore we would expect the partition function $Z_N$ to be
\begin{equation*}
Z_N=\mathcal{Z}^2_{N,A}\prod_{i=1}^{N-1}\mathcal{Z}^2_{N,B_i},
\end{equation*}
where $\mathcal{Z}^2_{N,A}$ and $\mathcal{Z}^2_{N,B_i}$ are just the second moment of the single interval density matrix. To see that we get the above decomposition, consider the tensor diagram for the case of two intervals $A$ and $B$ as shown in Figure \ref{fig:4iv}. The extension to arbitrary intervals in the faraway limit is straightforward. It was numerically checked in \cite{Coser:2013qda} that in this limit the theta function in the quantum part goes to unity, similarly, we checked numerically that $\Theta\left(0|T_N\right)$ also goes to unity. Finally, this also implies that the non-universal constant for $Z_{N}$ is $c^2_{N,2}$, where $c_{N,2}$ is just the non-universal constant for the $n=2$ R\'enyi entropy for the same set-up. We have the following expression for the total partition function $Z_{N}$,
\begin{equation} \label{eq4.3.1}
\begin{split}
Z_{N}=c^2_{N,2}\left|\frac{\prod_{j>i=2}^{N}(u_j-u_i)(v_j-v_i)}{(v_1-u_1)\prod_{j,i=2}^{N}(v_i-u_j)}\right|^{\frac{1}{2}}&  \left|\frac{\prod_{j=2}^{N}(u_j-u_1)(v_j-v_1)}{\prod_{j=2}^{N}(v_j-u_1)(u_j-v_1)}\right|^{\frac{1}{4}}\\
&\hspace{0.7in}\frac{\Theta\left(0|KT_N\right)\Theta\left(0|T_N/K\right)}{\Theta(0|\tau_{N-1})^2\Theta(0|\Omega_{N})^2}
\end{split}
\end{equation}
The $2$-R\'enyi CCNR negativity is just the logarithm of $Z_{N}$. As in the case of R\'enyi entropy and R\'enyi negativity, here too we manifestly have an invariance under $K\to 1/K$. The reflected R\'enyi entropy is given by
\begin{equation}
S_{2,2}(AA)=-\log\left(\frac{Z_{N}}{\mathcal{Z}_{N,2}^2}\right),
\end{equation}
where $\mathcal{Z}_{N,2}$ is the partition function obtained for $2$-R\'enyi entropy. We see that the ratio $Z_{N}/\mathcal{Z}_{N,2}^2$ is independent of any non-universal constants and hence the reflected entropy $S_{2,2}(AA)$ is a universal quantity. 

\subsubsection{Recovering the $N=2$ result}
Let us now consider the $N=2$ case and show that it matches with the known results in the literature. We have from eq.\eqref{eq4.3.1},
\begin{equation} \label{eq4.3.2}
Z_{2}=c^2_{2,2}\frac{(u_2-u_1)^{\frac{1}{4}}(v_2-v_1)^{\frac{1}{4}}}{(v_2-u_2)^{\frac{1}{2}}(v_1-u_1)^{\frac{1}{2}}}\frac{\vartheta_{3}\left(2\Omega_2 K\right)\vartheta_{3}\left(2\Omega_2/K\right)}{\vartheta_{3}\left(\Omega_2\right)^2}.
\end{equation}
We will proceed by substituting the moduli $\Omega_N$ of the Riemann surface $\Sigma_N$ with the moduli $2\Omega_2$ of the Riemann surface $\hat{\Sigma}_{2}$. Using the relation $y^{1/4}=\vartheta_4(\Omega_2)/\vartheta_3(\Omega_2)$, between the moduli $\Omega_2$ and the cross ratio $y$ (see eq.\eqref{eq4.1.5}-\eqref{eq4.1.6}), in the theta function relation
\begin{equation*}
\vartheta_3^2(\Omega_2)+\vartheta_3^2(\Omega_2)=2\vartheta_3^2(2\Omega_2),
\end{equation*}
we obtain the following relation between the moduli $\Omega_2$ and $2\Omega_2$,
\begin{equation} \label{eq4.3.3}
\vartheta_3^2(\Omega_2)=\frac{\vartheta_3^2(2\Omega_2)}{1+y^{\frac{1}{2}}}.
\end{equation}
Introducing the cross-ration $y'$, where $y'$ is related to $y$ by the following expression,
\begin{equation} \label{eq4.3.4}
y'=\frac{4y^{\frac{1}{2}}}{(1+y^{\frac{1}{2}})^2}.
\end{equation}
The cross ratio is related to the moduli $2\Omega_2$ by the following relation,
\begin{equation*}
2\Omega_2=i\frac{F\left(\frac{1}{2},\frac{1}{2};1;y'\right)}{F\left(\frac{1}{2},\frac{1}{2};1;1-y'\right)},
\end{equation*}
where $F$ is a hypergeometric function. Note that this is exactly the relation between $y$ and $\Omega_2$ as well. Thus we have a similar relation, $y'^{1/4}=\vartheta_4(2\Omega_2)/\vartheta_3(2\Omega_2)$ and $(1-y')^{1/4}=\vartheta_2(2\Omega_2)/\vartheta_3(2\Omega_2)$, using these relations and as well as eq.\eqref{eq4.3.4} in eq.\eqref{eq4.3.2}, we obtain the following,
\begin{equation} \label{eq4.3.5}
\vartheta_3^2(\Omega_2)=\frac{2\left(\vartheta_2\left(2\Omega_2\right)\vartheta_3\left(2\Omega_2\right)\vartheta_4\left(2\Omega_2\right)\right)^{\frac{2}{3}}}{y^{\frac{1}{12}}(1-y)^{\frac{1}{3}}}.
\end{equation}
Finally substituting this relation in eq.\eqref{eq4.3.2}, we obtain
\begin{equation*}
Z_{2}=\frac{\tilde{c}_2}{\left(\ell_1\ell_2(v_2-v_1)(v_2-u_1)(u_2-v_1)(u_2-u_1)\right)^{\frac{1}{6}}}\frac{\vartheta_{3}\left(2\Omega_2 K\right)\vartheta_{3}\left(2\Omega_2/K\right)}{\left(\vartheta_2\left(2\Omega_2\right)\vartheta_3\left(2\Omega_2\right)\vartheta_4\left(2\Omega_2\right)\right)^{\frac{2}{3}}},
\end{equation*}
where $\ell_1$ and $\ell_2$ are lengths of the intervals, and we absorbed any constants into the non-universal constant $\tilde{c}_2$. This is just the expression for $n=2$ R\'enyi CCNR negativity obtained in reference \cite{Yin:2022toc}.
\subsubsection{Dirac Fermions and Numerical checks}
The $2$-R\'enyi CCNR negativity for the Dirac fermion may be similarly obtained. It is known that the partition function of the Dirac fermion matches with that of the self-dual compact boson (i.e. $K=1$) when the Riemann period matrix $\Omega$ is completely imaginary \cite{Headrick:2012fk}, this is indeed the case in the present calculations. Hence, the twist field correlation for Dirac fermion is given by
\begin{equation} \label{eq4.3.6}
\begin{split}
Z_{N}=c^2_{N,2}\left|\frac{\prod_{j>i=2}^{N}(u_j-u_i)(v_j-v_i)}{(v_1-u_1)\prod_{j,i=2}^{N}(v_i-u_j)}\right|^{\frac{1}{2}}&  \left|\frac{\prod_{j=2}^{N}(u_j-u_1)(v_j-v_1)}{\prod_{j=2}^{N}(v_j-u_1)(u_j-v_1)}\right|^{\frac{1}{4}}\\
&\hspace{0.7in}\frac{\Theta\left(0|T_N\right)^2}{\Theta(0|\tau_{N-1})^2\Theta(0|\Omega_{N})^2}.
\end{split}
\end{equation}
We mention that the self-dual compact boson is not dual to the Dirac fermion, the known duality is between the $Z_2$-gauged modular invariant Dirac fermion and the compact boson at Bose-Fermi duality radius. The reflected entropy is given by
\begin{equation} \label{eq4.3.7}
S_{2,2}=-\log\left(\left|\frac{\prod_{j=2}^{N}(u_j-u_1)(v_j-v_1)}{\prod_{j=2}^{N}(v_j-u_1)(u_j-v_1)}\right|^{-\frac{1}{4}}\frac{\Theta\left(0|T_N\right)^2}{\Theta(0|\tau_{N-1})^2\Theta(0|\Omega_{N})^2}\right).
\end{equation} 
This result is matched against the tight-binding model for the case of three disjoint intervals, see Figures \ref{fig:4v}-\ref{fig:4vi} for the plots. The numerical model is given in the appendix \ref{C}. Since this ratio is universal there are no fitting parameters in these plots. We see from these plots that there is a very good match between the numerical results and the analytical predictions.
\begin{figure}
\centering % \begin{center}/\end{center} takes some additional vertical space
%\includegraphics[width=.5\textwidth]{tikz1.pdf}
% "\includegraphics" is very powerful; the graphicx package is already loaded
\includegraphics[width=1\textwidth]{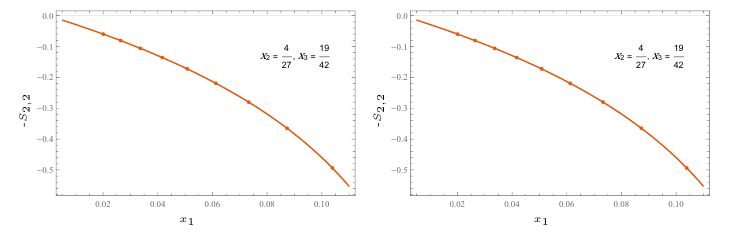}
\caption{\label{fig:4v} Reflected entropy plot for 3 Disjoint intervals: $-S_{2,2}$ is plotted against the cross-ratio $x_1$, with $x_2$ and $x_3$ fixed, see eq.\eqref{eq4.4}. The continuous lines are analytical plots for Dirac Fermions, while the plot points are numerical evaluations for the tight-binding model.}
\end{figure}
\begin{figure}
\centering % \begin{center}/\end{center} takes some additional vertical space
%\includegraphics[width=.5\textwidth]{tikz1.pdf}
% "\includegraphics" is very powerful; the graphicx package is already loaded
\includegraphics[width=1\textwidth]{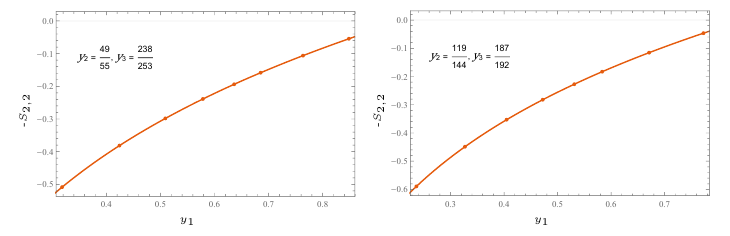}
\caption{\label{fig:4vi} Reflected entropy plot for three disjoint intervals: $-S_{2,2}$ is plotted against the cross-ratio $y_1$, with $y_2$ and $y_3$ fixed, see eq.\eqref{eq4.1.6}. The continuous lines are analytical plots for Dirac Fermions, while the plot points are numerical evaluations for the tight-binding model.}
\end{figure}
\subsection{Different Realignment} \label{section4.4}
We can also study the entanglement of another interval instead of $A$, with the union of the remaining intervals. In this case, we define the realignment matrix with respect to the desired interval. Let us generalise our techniques to this case as well. Consider the re-alignment with respect to the second interval as an example. We introduce the map $\tilde{w}$, mapping $u_2\to-\infty$, $v_2\to 0$, and $v_1\to 1$, we have
\begin{equation} \label{eq4.4.1}
\tilde{w}=\frac{(z-v_2)(u_2-v_1)}{(z-u_2)(v_2-v_1)}.
\end{equation}
The remaining branch points have the following mappings,
\begin{equation} \label{eq4.4.2}
\tilde{y}_{2j-1}= \frac{(u_{j+2}-v_2)(u_2-v_1)}{(u_{j+2}-u_2)(v_2-v_1)},\qquad \tilde{y}_{2j}= \frac{(v_{j+2}-v_2)(u_2-v_1)}{(v_{j+2}-u_2)(v_2-v_1)},
\end{equation}
where $j\in\{1,2,\ldots,N-3\}$ and $u_1\to \tilde{y}_{2N-3}$. The branch points lie on the real line of the $\tilde{w}$ plane and have the following order, $0<\tilde{y}_1<\tilde{y}_2\ldots<\tilde{y}_{2N-3}<1$. We then make a similar choice of the homology basis on the cut-$\title{w}$ surface as in Figure \ref{fig:4iii}. The remaining calculations follows exactly as in subsections \ref{section4.3}-\ref{section4.4}, with $y_i$ replaced with $\tilde{y}_i$. Notice that this is in general a different homology basis on the cut-$z$ surface, and with this choice, as is easily seen on the $\tilde{w}$ surface, one of the $a$ loops encircles only the second interval. So we are just considering a double cover of $\Sigma_N$ for the case where we have twisted boundary condition along the $b$ loop intersecting this $a$ loop.
\begin{figure}
\includegraphics[width=1\textwidth]{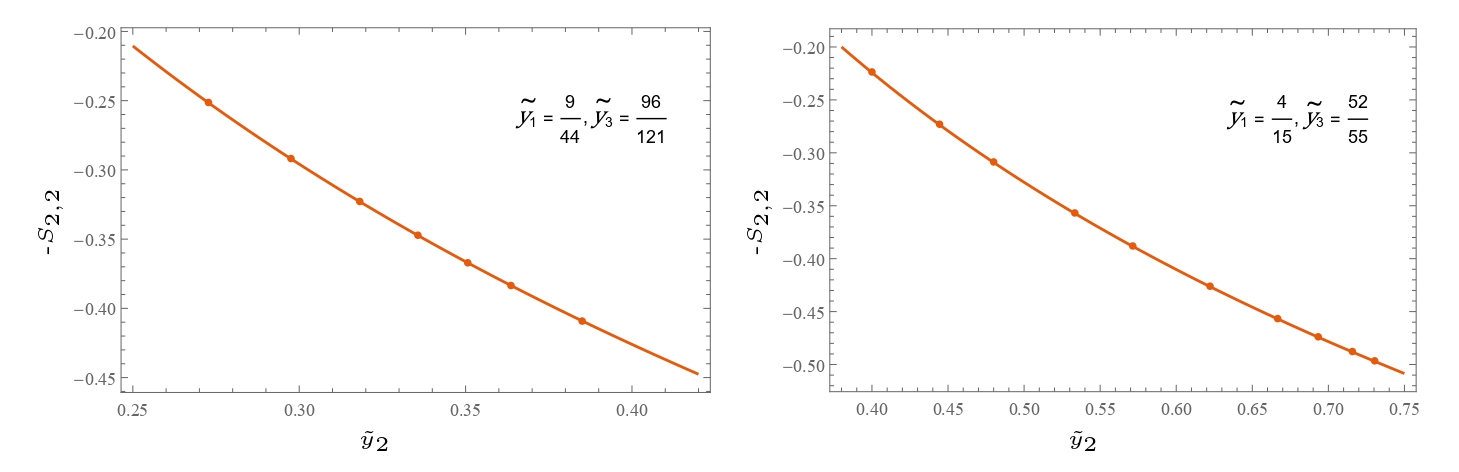}
\caption{\label{fig:4vii} Reflected entropy plot for Re-alignment with respect to the second interval ($3$ interval case): $-S_{2,2}$ is plotted against the cross-ratio $\tilde{y}_2$, with $\tilde{y}_1$ and $\tilde{y}_3$ fixed, see eq.\eqref{eq4.4.2}. The continuous lines are analytical plots for Dirac Fermions, while the plot points are numerical evaluations for the tight-binding model.}
\end{figure}

We have plotted the reflected entropy for three intervals with the above construction in Figure \ref{fig:4vii}. Here too we have plotted analytical results for the Dirac fermion against the numerical results of the tight-binding model, see appendix \ref{C}. We mention again that we don't have any free parameters in these plots, and similar to earlier plots, we have a very good match between the numerical evaluations and analytical predictions. 

\section{Conclusion} \label{section5}
In this work, we studied mixed-state entanglement in the two-dimensional compact boson conformal field theory (CFT) at arbitrary compactification radius, focusing on a configuration of $N$ disjoint intervals. We examined the entanglement between a single interval and the union of the remaining $N-1$ intervals by evaluating the $2$-Rényi computable cross norm (CCNR) negativity for this multi-interval set up.

We employed the replica trick and twist fields methods to compute the $2$-R\'enyi CCNR negativity. The correlation function of the relevant twist fields can be written as a product of a quantum part and a classical part. The quantum part was determined by diagonalising the twist fields and calculating the quantum component of the correlation functions for the diagonalised twist fields. The classical part was evaluated directly by computing the Riemann period matrix of the associated Riemann surface $\hat{\Sigma}$ in the replica trick. The logarithm of the twist fields correlation function gives $2$-R\'enyi CCNR negativity. The expression for $2$-R\'enyi CCNR negativity was given in terms of the period matrices of three Riemann surfaces. One of these surfaces was $\hat{\Sigma}$, the second was the Riemann surface for $2$-R\'enyi entropy of $N$-disjoint intervals, and the last one was the Riemann surface for $2$-R\'enyi entropy of $(N-1)$-disjoint intervals compromising one of the subsystems. We then obtained the reflected entropy related to the $2$-R\'enyi CCNR negativity. These reflected entropies are universal quantities, that is, we don't need a non-universal constant to compare the CFT results with the lattice models. We also extended our results to massless Dirac fermions. Finally, we checked our results against the tight-binding model and found a very good agreement between the analytical and numerical results.

There are several future directions of research worth pursuing. It will be interesting to extend these results to all integer values of the R\'enyi index. Since the Riemann surfaces associated with these calculations do not possess a $Z_n$-symmetry, as opposed to R\'enyi entropy and R\'enyi negativity, it would be very interesting to see the properties of R\'enyi CCNR negativities for multiple-disjoint interval settings. Another research direction would be to compute the $(m,n)$-R\'enyi CCNR negativity, introduced in \cite{Berthiere:2023gkx}, for the compact boson CFT, as similar features will also be present in these calculations even for two disjoint interval settings. Finally, the symmetry resolution of CCNR negativity for disjoint intervals is still an open problem for compact boson. The symmetry-resolved CCNR negativity of fermions and bosons has been studied in \cite{Berthiere:2023gkx} for adjacent intervals, and for massless Dirac fermions it has been studied in \cite{Bruno:2023tez} for two disjoint intervals. 
\appendix
\section{Riemann Surfaces} \label{A}
\begin{figure}
\centering % \begin{center}/\end{center} takes some additional vertical space
%\includegraphics[width=.5\textwidth]{tikz1.pdf}
% "\includegraphics" is very powerful; the graphicx package is already loaded
\includegraphics[width=0.95\textwidth]{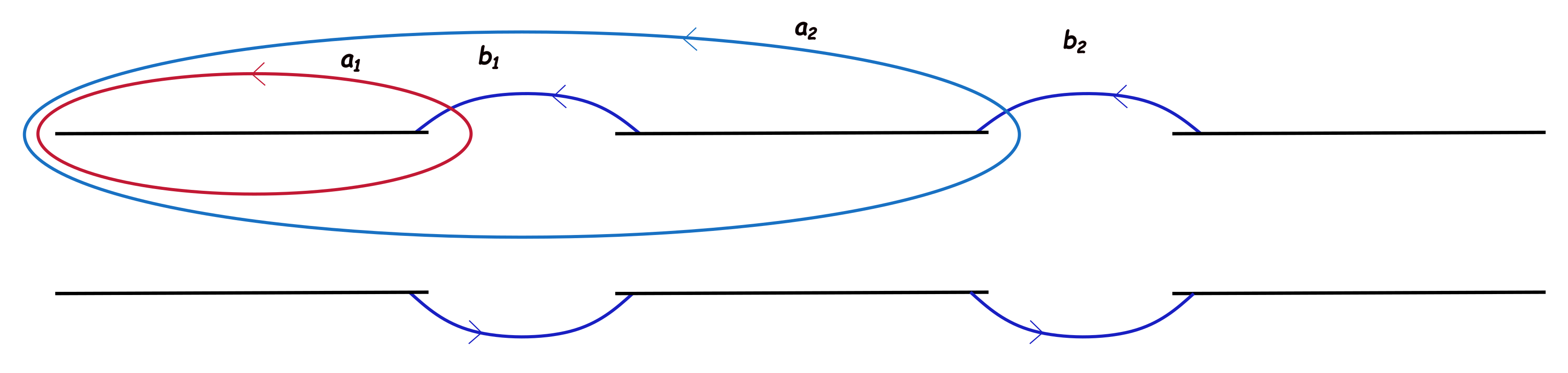}
\caption{\label{fig:Ai} Homology basis (N=3): Homology basis for $\Sigma_3$ is shown above, the horizontal lines represent branch cuts.}
\end{figure}

In this appendix, we review some concepts in the theory of Riemann surface relevant to the present work. We refer the reader to \cite{fay2006theta, mumford2007tata, Alvarez-Gaume:1987xem, Alvarez-Gaume:1986bwm} for a detailed review of Riemann surfaces. Specifically, we introduce homology basis, holomorphic differentials (also referred to as Abelian differentials) of the first kind and the Riemann period matrix for a given Riemann surface. In our discussions we will consider the Riemann surface $\tilde{\Sigma}_N$, parametrised by the following curve
\begin{equation*} 
u^2=\prod_{j=0}^{2N-2}(z-x_j)^{\frac{1}{2}},
\end{equation*}
this is the Riemann surface that we consider in Section \ref{section4} for $n=2$ R\'enyi entropy of multiple disjoint intervals, see discussion around eq.\eqref{eq4.7}. These results can be easily generalised to other cases, and we will refer to these results for our main computations. It follows from eq.\eqref{eq4.4}, $x_0=0<x_1<x_2\cdots<x_{2N-2}=1$. We consider the homology basis given by the $a$ and $b$ loops, that is the basis of non-contractible loops on the Riemann surface, as shown in Figure \ref{fig:Ai} for the specific case of $N=3$. The basis, irrespective of the Riemann surface, satisfies the following intersection rules
\begin{align}
a_i\circ a_j=b_i\circ b_j=0 \label{eqA.1}\\
a_i\circ b_j=-b_i\circ a_j=\delta_{i,j} \label{eqA.2}
\end{align}
The $a_i$ loops run counter-clockwise and encircle the first $i$ intervals of the branch cuts on the principal Riemann sheet. The $b_j$ loops run from the $(j+1)^{\text{th}}$ cut to the $j^{th}$ cut on the principle sheet and passes through the $j^{th}$ cut and meets itself after passing through the $(j+1)^{\text{th}}$ on the second sheet. This choice of $a$ and $b$ loops indeed satisfies the relations \eqref{eqA.1}-\eqref{eqA.2}.

Let us now pick the basis of the Abelian differentials of the first kind $\mathrm{d}w_i$, where $i\in\{1,2\cdots N-1\}$, to be the conventional one. They are given by
\begin{equation} \label{eqA.3}
\mathrm{d}w_i(z)=\frac{z^{i-1}}{u(z)}\mathrm{d}z.
\end{equation}
These Abelian differentials define the following $(N-1)\times(N-1)$ matrices $A$ and $B$,
\begin{equation} \label{eqA.4}
A_{i,j}=\oint_{a_j}\mathrm{d} w_i(z), \qquad B_{i,j}=\oint_{b_j}\mathrm{d} w_i(z).
\end{equation}
These contour integrals may be computed using the following set of integrals,
\begin{equation} \label{eqA.5}
\mathcal{I}^i_{x_{2j},x_{2j+1}}=\int_{x_{2j}}^{x_{2j+1}}\mathrm{d}w_{i}(z),
\end{equation}
which are just Lauricella functions \cite{Coser:2013qda}. We can now introduce normalised holomorphic differentials $\mathrm{d}\nu_i(z)$,
\begin{equation} \label{eqA.6}
\mathrm{d}\nu_i(z)=A^{-1}_{i,j}\mathrm{d}w_{j}(z),
\end{equation}
so that they satisfy $\oint_{a_i}\mathrm{d}\nu_j(z)=\delta_{i,j}$. Finally, using the normalised holomorphic differentials we define the Riemann period matrix $\Pi$,
\begin{equation} \label{eqA.7}
\Pi_{i,j}=\oint_{b_j}\mathrm{d}z\nu_{i}(z).
\end{equation}
Note that the period matrix is independent of the choice of the basis of the Abelian differentials of the first kind and are given in the homology basis $a$ and $b$. As mentioned in the main text, $\Pi$ is a symmetric matrix with a positive definite imaginary part. In fact, in the present case, that is for Riemann surfaces given by eq.\eqref{eq4.7}, the real part of the period matrix vanishes and hence it is purely imaginary. 
\section{Riemann Surface $\hat{\Sigma}_N$} \label{B}
In this appendix, we find the Abelian differentials of the first kind on the Riemann surface $\hat{\Sigma}_N$. We also give the matrices $A$ and $B$ by computing the contour integral of these differentials along the loops $\hat{a}$ and $\hat{b}$  

As mentioned in subsection \ref{section4.2}, the first $(N-1)$ of the Abelian differentials of the first kind are constructed by lifting the normalised holomorphic differentials $\mathrm{d}\nu_j$ on $\Sigma_N$, we substitute $w=\zeta^2$ in the normalised differentials to obtain
\begin{equation} \label{eqB.1}
\mathrm{d}w_i=\left(A_N^{(-1)}\right)_{ij}\zeta^{2(j-1)}\prod_{k=1}^{2N-2}\left(\zeta-\sqrt{y_k}\right)^{-\frac{1}{2}}\left(\zeta-\sqrt{y_i}\right)^{-\frac{1}{2}} \mathrm{d}\zeta,
\end{equation} 
where we have absorbed any constant into the definition of $\mathrm{d}w_i$, and the matrix $A_N$ is evaluated on $\Sigma_N$ as discussed in appendix \ref{A}, specifically see eq.\eqref{eqA.4}. The Prym differentials may be constructed from the Abelian differentials of the first kind on $\Sigma'_{N-1}$, given by eq.\eqref{eq4.1.16}-\eqref{eq4.1.17}, for reference, we reproduce them here
\begin{equation*}
\mathrm{d}\mu'_j=w^{j-1}\prod_{i=1}^{2N-2}(w-y_k)^{-\frac{1}{2}}.
\end{equation*}
These differentials have a zero period around the loops ${a}_1$ and ${b}_1$ on $\Sigma_N$. To see this, we note that loop $a_1$ is contractible on $\Sigma'_{N-1}$, and loop $b_1$ encloses only one branch point, in fact, $b_1$ is not a closed loop on $\Sigma'_{N-1}$. Finally, we also note that $\mathrm{d}\mu'_j$ are odd when taken around the loop $b_1$ on $\Sigma_N$. This implies that the normalised holomorphic differentials on $\Sigma'_{N-1}$ satisfy the properties of Prym differentials on $\Sigma_N$, as discussed in the main text. Therefore, the remaining $N-2$ Abelian differentials of the first kind or the Prym differentials on $\hat{\Sigma}_N$ are given by,
\begin{equation} \label{eqB.2}
\mathrm{d}w_{2N-3-i}=\left(A_{N-1}^{'(-1)}\right)_{ij}\zeta^{2j-1}\prod_{k=1}^{2N-2}\left(\zeta-\sqrt{y_k}\right)^{-\frac{1}{2}}\left(\zeta-\sqrt{y_k}\right)^{-\frac{1}{2}}\mathrm{d}\zeta,
\end{equation} 
where $I\in\{1,\ldots,N-2\}$ and as before we have absorbed any constant into $\mathrm{d}w_{2N-3-j}$. The matrix $A_{N-1}'$ is just the $A$ matrix of eq.\eqref{eqA.4} on $\Sigma'_{N-1}$.

As mentioned in the main text, the conformal automorphism on $\hat{\Sigma}_N$ is a reflection through $\zeta=0$ followed by a sheet interchange. Now, to evaluate the contour integrals of $\mathrm{d}w_j$ around the homology basis $\hat{a}$ and $\hat{b}$, see Figure \ref{fig:4iii}, we must carefully pick the correct branches for the integrand. We choose the principal branch for the integrand to be below the cuts for all the branch cuts to the left of the origin on the first sheet. Now we may evaluate the matrices $A$ and $B$ by using the following integrals,
\begin{equation} \label{eqB.3}
\begin{split}
I^j_{\sqrt{y_i},\sqrt{y_{j+i}}}&=\int_{\sqrt{y_i}}^{\sqrt{y_{i+1}}}\mathrm{d}\zeta\zeta^{2(j-1)}\prod_{k=1}^{2N-2}\left(\zeta-\sqrt{y_k}\right)^{-\frac{1}{2}}\left(\zeta-\sqrt{y_k}\right)^{-\frac{1}{2}}\\
&=\frac{1}{2}\int_{y_i}^{y_{i+1}}\mathrm{d}zz^{j-1}\prod_{k=0}^{2N-1}(z-y_k)^{-\frac{1}{2}},
\end{split}
\end{equation}
where $j\in\{1,2\cdots N-1\}$. Similarly for the remaining $\mathrm{d}w_j$ we have,
\begin{equation} \label{eqB.4}
I^{2N-3-j}_{\sqrt{y_i},\sqrt{y_{j+i}}}=\frac{1}{2}\int_{y_i}^{y_{i+1}}\mathrm{d}zz^{j-1}\prod_{k=1}^{2N-1}(z-y_k)^{-\frac{1}{2}}.
\end{equation}
We also note that under the conformal automorphism $\alpha$, the differentials $\mathrm{d}w_j$ are even when $j\in\{1,\ldots,N-1\}$ and odd when $j\in\{N,\ldots,2N-3\}$, this in accordance with eq.\eqref{eq4.2.5}-\eqref{eq4.2.6}. The  matrix $B$ maybe written in terms of the Riemann period matrices $\Omega_N$ and $\tau_{N-1}$ of Riemann surfaces $\Sigma_N$ and $\Sigma'_{N-1}$ respectively. These matrices are found to be
\begin{equation} \label{eqB.5}
{A}_{i j}=\left\{\begin{array}{rl}
\frac{1}{2} & \hspace{0.3in} j=i ;\hspace{0.05in} i \in\{1, \ldots, N-1\} \\
\frac{1}{2} & \hspace{0.3in} j=2 N-1-i ;\hspace{0.05in} i \in\{2, \ldots, N-1\} \\
\frac{1}{2} & \hspace{0.3in} j=i-N+2 ;\hspace{0.05in} i \in\{N, \cdots, 2 N-3\} \\
-\frac{1}{2} & \hspace{0.3in} j=3 N-3-i ;\hspace{0.05in} i \in\{N, \ldots, 2 N-3\} \\
0 & \hspace{0.3in} \text { otherwise, }
\end{array}\right.
\end{equation}
\begin{equation} \label{eqB.6}
{B}_{i j}=\left\{\begin{array}{rl}
\Omega_{i,1} & \hspace{0.3in}  i \in\{1, \ldots, N-1\} \\
\frac{1}{2}\Omega_{i,j} & \hspace{0.3in} i,j \in\{2, \ldots, N-1\} \\
\frac{1}{2}\Omega_{i,2N-1-j} & \hspace{0.3in} i \in\{2, \ldots, N-1\},\text{ }j\in\{N,\ldots, 2N-3\} \\
\frac{1}{2}\tau_{i-N+1,j-1} & \hspace{0.3in} i \in\{N,\ldots, 2N-3\},\text{ }j\in\{2, \ldots, N-1\} \\
-\frac{1}{2}\tau_{i-N+1,2N-2-j} & \hspace{0.3in} i \in\{N,\ldots, 2N-3\},\text{ }j\in\{N,\ldots, 2N-3\} \\
0 & \hspace{0.3in} \text { otherwise. }
\end{array}\right. ,
\end{equation}
where we have dropped the subscript on $\Omega$ and $\tau$ above for brevity. Note we did not normalise the differentials $\mathrm{d}\hat{w}$ according to eq.\eqref{eq4.2.3}-\eqref{eq4.2.4} and so we get factors of $\frac{1}{2}$ in eq.\eqref{eqB.5}-\eqref{eqB.6}, but this is really of no consequence. The inverse of matrix $A$ is given by,
\begin{equation} \label{eqB.7}
{A^{(-1)}}_{i j}=\left\{\begin{array}{rl}
2 & \hspace{0.3in} i=j=1  \\
1 & \hspace{0.3in} i=j ;\hspace{0.05in} i \in\{1, \ldots, N-1\} \\
1 & \hspace{0.3in} i=2N-1-j ;\hspace{0.05in} j\in\{2, \ldots, N-1\} \\
1 & \hspace{0.3in} i=j-N+2 ;\hspace{0.05in} j \in\{N, \ldots, 2 N-3\} \\
-1 & \hspace{0.3in} i=3N-3-j ;\hspace{0.05in} j \in\{N, \ldots, 2 N-3\} \\
0 & \hspace{0.3in} \text { otherwise. }
\end{array}\right.
\end{equation}
We have used these results in the main text to obtain the period matrix $T$ for $\hat{\Sigma}_N$, see eq.\eqref{eq4.2.7}. 
\section{Numerical Model} \label{C}
In this section, we give the numerical evaluation of R\'enyi CCNR negativity and reflected entropies for the tight-binding model. The results for reflected entropies are used in the plots of Figures \ref{fig:4v}, \ref{fig:4vi} and \ref{fig:4vii}. This method for reflected entropy was first developed in \cite{Bueno:2020vnx, Bueno:2020fle} and later was generalised to R\'enyi CCNR negativities and $(m,n)$-R\'enyi reflected entropies in \cite{Berthiere:2023gkx} for both fermions and bosons.

Tight-binding model has the Hamiltonian $H=-\sum_i \hat{c}^{\dagger}_{i+1}\hat{c}_{i}+\hat{c}^{\dagger}_{i}\hat{c}_{i+1}$. The fermionic operators $\hat{c}_i$ satisfies the following commutation relations $\{\hat{c}_i,\hat{c}^{\dagger}_j\}=\delta_{i,j}$. The correlation matrix $C_{ij}=\left\langle\hat{c}^{\dagger}_i\hat{c}_j \right\rangle$ for the tight binding model is given by
\begin{equation} \label{eqD.1}
C_{ij}=\frac{\sin{\left((i-j)\pi/2\right)}}{(i-j)\pi}.
\end{equation}
The reduced density matrix for the subsystem $A\cup B$ may be given by the diagonalised Gaussian state,
\begin{equation} \label{eqD.2}
\rho_{AB}=\prod_{k}\frac{e^{-\epsilon_k a_k^{\dagger}a_k}}{1+e^{-\epsilon_k}},
\end{equation}
where the fermion operators $a_k$ are related to the operators $c_i$, where $i\in\{A\cup B\}$, via an unitary transformation. The spectrum $\epsilon_k$ of the Gaussian state is related to the eigenvalues $\frac{1+\nu_k}{2}$ of the correlation matrix $C_{AB}$ via \cite{Peschel:2002yqj},
\begin{equation} \label{eqD.3}
\frac{1}{1+e^{\epsilon_k}}=\frac{1+\nu_k}{2},
\end{equation}
where $C_{AB}$ is the restriction of the correlation matrix $C$ to the subsystem $A\cup B$. The R\'enyi entropies for the reduced density matrix $\rho_{AB}$ is then given by
\begin{equation} \label{eqD.4}
S_{n}(AB)=\frac{1}{1-n}\mathrm{Tr}\log(C_{AB}^n+(1-C_{AB})^n).
\end{equation}
To generalise this formalism to ($m,n$)-reflected entropies and CCNR R\'enyi negativities, we consider the Choi-Jamiolkowski isomorphism for $\rho_{AB}^m$,
\begin{equation}
|\Omega_m\rangle=\frac{1}{\sqrt{\mathrm{Tr}\rho_{AB}^m}}\prod_{k}\frac{\left(1+e^{-\frac{m\epsilon_k}{2}}a_k^{\dagger} \tilde{a}^{\dagger}_k\right)}{\left(1+e^{-\epsilon_k}\right)^{\frac{m}{2}}}|0\rangle|\tilde{0}\rangle.
\end{equation}
where the operators $a_k$ and $\tilde{a}_k$ belong to the either copy of the hilbert space $\mathcal{H}_{AB}$. The correlation matrix for these operators is given by
\begin{equation}
\langle\Omega_m|\left(\begin{array}{c} a_k^{\dagger}\\ \tilde{a}_k \end{array}\right)\left(\begin{array}{cc} a_{k'} & \tilde{a}_{k'}^{\dagger} \end{array}\right)|\Omega_m\rangle=\delta_{kk'}\frac{\left(1+e^{\epsilon_k}\right)^m}{1+e^{m\epsilon_k}} \left[\begin{array}{cc}
\frac{1}{\left(1+e^{-\epsilon_k}\right)^m} & \frac{e^{-\frac{m\epsilon_k}{2}}}{\left(1+e^{-\epsilon_k}\right)^m} \\ \frac{e^{-\frac{m\epsilon_k}{2}}}{\left(1+e^{-\epsilon_k}\right)^m} &\frac{e^{-m\epsilon_k}}{\left(1+e^{-\epsilon_k}\right)^m}
\end{array}\right].
\end{equation}
Let's denote this correlation matrix by $\tilde{C}^{(m)}$, in the (doubled) spatial basis, this correlation matrix is given by,
\begin{equation}
\tilde{C}^{(m)}= \frac{1}{{C_{AB}^m+(1-C_{AB})^m}}\left[\begin{array}{cc}
{C_{AB}^m} & {C_{AB}^\frac{m}{2}(1-C_{AB})^\frac{m}{2}} \\ {C_{AB}^\frac{m}{2}(1-C_{AB})^\frac{m}{2}} & {(1-C_{AB})^m}
\end{array}\right].
\end{equation}
We now assume that the reflected reduced density matrix ${\rho}_{AA}$ is also given by a diagonalised Gaussian state,
\begin{equation}
{\rho}_{AA}=\prod_{k}\frac{e^{-\tilde{\epsilon_k} b_k^{\dagger}b_k}}{1+e^{-\tilde{\epsilon_k}}},
\end{equation}  
where the operators $b_k$ are related to the operators $c_i$ and $\tilde{c}_i^{\dagger}$, where $i\in\{A\}$, by a unitary transformation. Then from eq.\eqref{eqD.2}-\eqref{eqD.4}, we conclude that the $(m,n)$-reflected R\'enyi entropy is given by,
\begin{equation}
S_{m,n}=\frac{1}{1-n}\mathrm{Tr}\log\left(\left(\tilde{C}^{(m)}_A\right)^n+\left(1-\tilde{C}^{(m)}_A\right)^n\right),
\end{equation}
where $\tilde{C}^{(m)}_A$ is the restriction of $\tilde{C}^{(m)}$ to the degrees of freedom in the doubled subspace $A$. We are interested in the case $m=n=2$ for the numerical plots in Figures \ref{fig:4v}, \ref{fig:4vi} and \ref{fig:4vii}. Finally, the CCNR negativity is given by using the relation in eq.\eqref{eq2.7}.
\acknowledgments 
HG is supported by the Prime Minister’s Research Fellowship offered by the Ministry of
Education, Govt. of India. HG would like to thank Urjit A. Yajnik for their encouragement, helpful discussions, and comments on the manuscript. HG also thanks J\'er\^ome Dubail for helpful discussions. We would also like to thank the anonymous referee for their insightful comments on the manuscript.
\bibliographystyle{unsrt}
\bibliography{BibtexCCNR.bib}

\begin{thebibliography}{100}

\bibitem{Ryu:2006bv}
Shinsei Ryu and Tadashi Takayanagi.
\newblock {Holographic derivation of entanglement entropy from AdS/CFT}.
\newblock {\em Phys. Rev. Lett.}, 96:181602, 2006.

\bibitem{Nishioka:2009un}
Tatsuma Nishioka, Shinsei Ryu, and Tadashi Takayanagi.
\newblock {Holographic Entanglement Entropy: An Overview}.
\newblock {\em J. Phys. A}, 42:504008, 2009.

\bibitem{Solodukhin:2011gn}
Sergey~N. Solodukhin.
\newblock {Entanglement entropy of black holes}.
\newblock {\em Living Rev. Rel.}, 14:8, 2011.

\bibitem{Nielsen_Chuang_2010}
Michael~A. Nielsen and Isaac~L. Chuang.
\newblock {\em Quantum Computation and Quantum Information: 10th Anniversary
  Edition}.
\newblock Cambridge University Press, 2010.

\bibitem{Amico:2007ag}
Luigi Amico, Rosario Fazio, Andreas Osterloh, and Vlatko Vedral.
\newblock {Entanglement in many-body systems}.
\newblock {\em Rev. Mod. Phys.}, 80:517--576, 2008.

\bibitem{Osterloh:2002sym}
A.~Osterloh, Luigi Amico, G.~Falci, and Rosario Fazio.
\newblock {Scaling of entanglement close to a quantum phase transition}.
\newblock {\em Nature}, 416(6881):608--610, 2002.

\bibitem{Vidal:2002rm}
G.~Vidal, J.~I. Latorre, E.~Rico, and A.~Kitaev.
\newblock {Entanglement in quantum critical phenomena}.
\newblock {\em Phys. Rev. Lett.}, 90:227902, 2003.

\bibitem{Calabrese:2004eu}
Pasquale Calabrese and John~L. Cardy.
\newblock {Entanglement entropy and quantum field theory}.
\newblock {\em J. Stat. Mech.}, 0406:P06002, 2004.

\bibitem{Peres:1996dw}
Asher Peres.
\newblock {Separability criterion for density matrices}.
\newblock {\em Phys. Rev. Lett.}, 77:1413--1415, 1996.

\bibitem{Horodecki:1996nc}
Michal Horodecki, Pawel Horodecki, and Ryszard Horodecki.
\newblock {On the necessary and sufficient conditions for separability of mixed
  quantum states}.
\newblock {\em Phys. Lett. A}, 223:1, 1996.

\bibitem{Rudolph:2005tpa}
Oliver Rudolph.
\newblock {Further Results on the Cross Norm Criterion for Separability}.
\newblock {\em Quant. Inf. Proc.}, 4(3):219--239, 2005.

\bibitem{Rudolph:2002qos}
Oliver Rudolph.
\newblock {Some properties of the computable cross-norm criterion for
  separability}.
\newblock {\em Phys. Rev. A}, 67(3):032312, 2003.

\bibitem{Chen:2003slc}
K.~Chen and L.~A. Wu.
\newblock {A matrix realignment method for recognizing entanglement}.
\newblock {\em Quant. Inf. Comput.}, 3(3):193--202, 2003.

\bibitem{Yin:2022toc}
Chao Yin and Zhenhuan Liu.
\newblock {Universal Entanglement and Correlation Measure in Two-Dimensional
  Conformal Field Theories}.
\newblock {\em Phys. Rev. Lett.}, 130(13):131601, 2023.

\bibitem{Vidal:2002zz}
G.~Vidal and R.~F. Werner.
\newblock {Computable measure of entanglement}.
\newblock {\em Phys. Rev. A}, 65:032314, 2002.

\bibitem{Plenio:2005cwa}
M.~B. Plenio.
\newblock {Logarithmic Negativity: A Full Entanglement Monotone That is not
  Convex}.
\newblock {\em Phys. Rev. Lett.}, 95:090503, 2005.

\bibitem{Berthiere:2023gkx}
Cl\'ement Berthiere and Gilles Parez.
\newblock {Reflected entropy and computable cross-norm negativity: Free
  theories and symmetry resolution}.
\newblock {\em Phys. Rev. D}, 108(5):054508, 2023.

\bibitem{Milekhin:2022zsy}
Alexey Milekhin, Pratik Rath, and Wayne Weng.
\newblock {Computable Cross Norm in Tensor Networks and Holography}.
\newblock 12 2022.

\bibitem{Dutta:2019gen}
Souvik Dutta and Thomas Faulkner.
\newblock {A canonical purification for the entanglement wedge cross-section}.
\newblock {\em JHEP}, 03:178, 2021.

\bibitem{Holzhey:1994we}
Christoph Holzhey, Finn Larsen, and Frank Wilczek.
\newblock {Geometric and renormalized entropy in conformal field theory}.
\newblock {\em Nucl. Phys. B}, 424:443--467, 1994.

\bibitem{Calabrese:2009qy}
Pasquale Calabrese and John Cardy.
\newblock {Entanglement entropy and conformal field theory}.
\newblock {\em J. Phys. A}, 42:504005, 2009.

\bibitem{Fradkin:2006mb}
Eduardo Fradkin and Joel~E. Moore.
\newblock {Entanglement entropy of 2D conformal quantum critical points:
  hearing the shape of a quantum drum}.
\newblock {\em Phys. Rev. Lett.}, 97:050404, 2006.

\bibitem{Cardy:2016fqc}
John Cardy and Erik Tonni.
\newblock {Entanglement hamiltonians in two-dimensional conformal field
  theory}.
\newblock {\em J. Stat. Mech.}, 1612(12):123103, 2016.

\bibitem{Alcaraz:2011tn}
Francisco~Castilho Alcaraz, Miguel~Ibanez Berganza, and German Sierra.
\newblock {Entanglement of low-energy excitations in Conformal Field Theory}.
\newblock {\em Phys. Rev. Lett.}, 106:201601, 2011.

\bibitem{Hsu:2008af}
Benjamin Hsu, Michael Mulligan, Eduardo Fradkin, and Eun-Ah Kim.
\newblock {Universal entanglement entropy in 2D conformal quantum critical
  points}.
\newblock {\em Phys. Rev. B}, 79:115421, 2009.

\bibitem{Calabrese:2014yza}
Pasquale Calabrese, John Cardy, and Erik Tonni.
\newblock {Finite temperature entanglement negativity in conformal field
  theory}.
\newblock {\em J. Phys. A}, 48(1):015006, 2015.

\bibitem{Shapourian:2016cqu}
Hassan Shapourian, Ken Shiozaki, and Shinsei Ryu.
\newblock {Partial time-reversal transformation and entanglement negativity in
  fermionic systems}.
\newblock {\em Phys. Rev. B}, 95(16):165101, 2017.

\bibitem{Calabrese:2005in}
Pasquale Calabrese and John~L. Cardy.
\newblock {Evolution of entanglement entropy in one-dimensional systems}.
\newblock {\em J. Stat. Mech.}, 0504:P04010, 2005.

\bibitem{Calabrese:2007mtj}
Pasquale Calabrese and John Cardy.
\newblock {Entanglement and correlation functions following a local quench: a
  conformal field theory approach}.
\newblock {\em J. Stat. Mech.}, 0710(10):P10004, 2007.

\bibitem{Wen:2015qwa}
Xueda Wen, Po-Yao Chang, and Shinsei Ryu.
\newblock {Entanglement negativity after a local quantum quench in conformal
  field theories}.
\newblock {\em Phys. Rev. B}, 92(7):075109, 2015.

\bibitem{Coser:2014gsa}
Andrea Coser, Erik Tonni, and Pasquale Calabrese.
\newblock {Entanglement negativity after a global quantum quench}.
\newblock {\em J. Stat. Mech.}, 1412(12):P12017, 2014.

\bibitem{Hoogeveen:2014bqa}
Marianne Hoogeveen and Benjamin Doyon.
\newblock {Entanglement negativity and entropy in non-equilibrium conformal
  field theory}.
\newblock {\em Nucl. Phys. B}, 898:78--112, 2015.

\bibitem{Cardy:2007mb}
J.~L. Cardy, O.~A. Castro-Alvaredo, and B.~Doyon.
\newblock {Form factors of branch-point twist fields in quantum integrable
  models and entanglement entropy}.
\newblock {\em J. Statist. Phys.}, 130:129--168, 2008.

\bibitem{Doyon:2008vu}
Benjamin Doyon.
\newblock {Bi-partite entanglement entropy in massive two-dimensional quantum
  field theory}.
\newblock {\em Phys. Rev. Lett.}, 102:031602, 2009.

\bibitem{Castro-Alvaredo:2009yqb}
Olalla~A. Castro-Alvaredo and Benjamin Doyon.
\newblock {Bi-partite entanglement entropy in massive 1+1-dimensional quantum
  field theories}.
\newblock {\em J. Phys. A}, 42:504006, 2009.

\bibitem{Bianchini:2015uea}
Davide Bianchini, Olalla~A. Castro-Alvaredo, and Benjamin Doyon.
\newblock {Entanglement Entropy of Non-Unitary Integrable Quantum Field
  Theory}.
\newblock {\em Nucl. Phys. B}, 896:835--880, 2015.

\bibitem{Blondeau-Fournier:2015yoa}
Olivier Blondeau-Fournier, Olalla~A. Castro-Alvaredo, and Benjamin Doyon.
\newblock {Universal scaling of the logarithmic negativity in massive quantum
  field theory}.
\newblock {\em J. Phys. A}, 49(12):125401, 2016.

\bibitem{Castro-Alvaredo:2018dja}
Olalla~A. Castro-Alvaredo, Cecilia De~Fazio, Benjamin Doyon, and Istv\'an~M.
  Sz\'ecs\'enyi.
\newblock {Entanglement Content of Quasiparticle Excitations}.
\newblock {\em Phys. Rev. Lett.}, 121(17):170602, 2018.

\bibitem{Castro-Alvaredo:2008fni}
Olalla~A. Castro-Alvaredo and Benjamin Doyon.
\newblock {Bi-partite entanglement entropy in massive QFT with a boundary: The
  Ising model}.
\newblock {\em J. Statist. Phys.}, 134:105--145, 2009.

\bibitem{zhou2006entanglement}
Huan-Qiang Zhou, Thomas Barthel, John~Ove Fj{\ae}restad, and Ulrich
  Schollw{\"o}ck.
\newblock Entanglement and boundary critical phenomena.
\newblock {\em Physical Review A—Atomic, Molecular, and Optical Physics},
  74(5):050305, 2006.

\bibitem{Cornfeld:2017tkz}
Eyal Cornfeld and Eran Sela.
\newblock {Entanglement entropy and boundary renormalization group flow: Exact
  results in the Ising universality class}.
\newblock {\em Phys. Rev. B}, 96(7):075153, 2017.

\bibitem{Estienne:2023ekf}
Benoit Estienne, Yacine Ikhlef, Andrei Rotaru, and Erik Tonni.
\newblock {Entanglement entropies of an interval for the massless scalar field
  in the presence of a boundary}.
\newblock {\em JHEP}, 05:236, 2024.

\bibitem{Capizzi:2022xdt}
Luca Capizzi, Sara Murciano, and Pasquale Calabrese.
\newblock {R\'enyi entropy and negativity for massless Dirac fermions at
  conformal interfaces and junctions}.
\newblock {\em JHEP}, 08:171, 2022.

\bibitem{Capizzi:2022uni}
Luca Capizzi, Sara Murciano, and Pasquale Calabrese.
\newblock {R\'enyi entropy and negativity for massless complex boson at
  conformal interfaces and junctions}.
\newblock {\em JHEP}, 11:105, 2022.

\bibitem{Gutperle:2017enx}
Michael Gutperle and John~D. Miller.
\newblock {Entanglement entropy at CFT junctions}.
\newblock {\em Phys. Rev. D}, 95(10):106008, 2017.

\bibitem{Ohmori:2014eia}
Kantaro Ohmori and Yuji Tachikawa.
\newblock {Physics at the entangling surface}.
\newblock {\em J. Stat. Mech.}, 1504:P04010, 2015.

\bibitem{Rogerson:2022yim}
David Rogerson, Frank Pollmann, and Ananda Roy.
\newblock {Entanglement entropy and negativity in the Ising model with
  defects}.
\newblock {\em JHEP}, 06:165, 2022.

\bibitem{Capizzi:2022igy}
Luca Capizzi and Viktor Eisler.
\newblock {Entanglement evolution after a global quench across a conformal
  defect}.
\newblock {\em SciPost Phys.}, 14(4):070, 2023.

\bibitem{Castro-Alvaredo:2024azg}
Olalla~A. Castro-Alvaredo and Luc\'\i{}a Santamar\'\i{}a-Sanz.
\newblock {Symmetry Resolved Measures in Quantum Field Theory: a Short Review}.
\newblock 3 2024.

\bibitem{Goldstein:2017bua}
Moshe Goldstein and Eran Sela.
\newblock {Symmetry-resolved entanglement in many-body systems}.
\newblock {\em Phys. Rev. Lett.}, 120(20):200602, 2018.

\bibitem{Cornfeld:2018wbg}
Eyal Cornfeld, Moshe Goldstein, and Eran Sela.
\newblock {Imbalance entanglement: Symmetry decomposition of negativity}.
\newblock {\em Phys. Rev. A}, 98(3):032302, 2018.

\bibitem{Xavier:2018kqb}
J.~C. Xavier, F.~C. Alcaraz, and G.~Sierra.
\newblock {Equipartition of the entanglement entropy}.
\newblock {\em Phys. Rev. B}, 98(4):041106, 2018.

\bibitem{Murciano:2021djk}
Sara Murciano, Riccarda Bonsignori, and Pasquale Calabrese.
\newblock {Symmetry decomposition of negativity of massless free fermions}.
\newblock {\em SciPost Phys.}, 10(5):111, 2021.

\bibitem{Jones:2022tgp}
Nick~G. Jones.
\newblock {Symmetry-Resolved Entanglement Entropy in Critical Free-Fermion
  Chains}.
\newblock {\em J. Statist. Phys.}, 188(3):28, 2022.

\bibitem{Murciano:2020vgh}
Sara Murciano, Giuseppe Di~Giulio, and Pasquale Calabrese.
\newblock {Entanglement and symmetry resolution in two dimensional free quantum
  field theories}.
\newblock {\em JHEP}, 08:073, 2020.

\bibitem{Foligno:2022ltu}
Alessandro Foligno, Sara Murciano, and Pasquale Calabrese.
\newblock {Entanglement resolution of free Dirac fermions on a torus}.
\newblock {\em JHEP}, 03:096, 2023.

\bibitem{Capizzi:2021kys}
Luca Capizzi, D\'avid~X. Horv\'ath, Pasquale Calabrese, and Olalla~A.
  Castro-Alvaredo.
\newblock {Entanglement of the 3-state Potts model via form factor bootstrap:
  total and symmetry resolved entropies}.
\newblock {\em JHEP}, 05:113, 2022.

\bibitem{Horvath:2021fks}
David~X. Horvath, Luca Capizzi, and Pasquale Calabrese.
\newblock {U(1) symmetry resolved entanglement in free 1+1 dimensional field
  theories via form factor bootstrap}.
\newblock {\em JHEP}, 05:197, 2021.

\bibitem{Parez:2020vsp}
Gilles Parez, Riccarda Bonsignori, and Pasquale Calabrese.
\newblock {Quasiparticle dynamics of symmetry-resolved entanglement after a
  quench: Examples of conformal field theories and free fermions}.
\newblock {\em Phys. Rev. B}, 103(4):L041104, 2021.

\bibitem{DiGiulio:2022jjd}
Giuseppe Di~Giulio, Ren\'e Meyer, Christian Northe, Henri Scheppach, and Suting
  Zhao.
\newblock {On the boundary conformal field theory approach to symmetry-resolved
  entanglement}.
\newblock {\em SciPost Phys. Core}, 6:049, 2023.

\bibitem{Kusuki:2023bsp}
Yuya Kusuki, Sara Murciano, Hirosi Ooguri, and Sridip Pal.
\newblock {Symmetry-resolved entanglement entropy, spectra \& boundary
  conformal field theory}.
\newblock {\em JHEP}, 11:216, 2023.

\bibitem{Calabrese:2021wvi}
Pasquale Calabrese, J\'er\^ome Dubail, and Sara Murciano.
\newblock {Symmetry-resolved entanglement entropy in Wess-Zumino-Witten
  models}.
\newblock {\em JHEP}, 10:067, 2021.

\bibitem{zhao2021symmetry}
Suting Zhao, Christian Northe, and Ren{\'e} Meyer.
\newblock Symmetry-resolved entanglement in ads3/cft2 coupled to u (1)
  chern-simons theory.
\newblock {\em Journal of High Energy Physics}, 2021(7):1--38, 2021.

\bibitem{weisenberger2021symmetry}
Konstantin Weisenberger, Suting Zhao, Christian Northe, and Ren{\'e} Meyer.
\newblock Symmetry-resolved entanglement for excited states and two entangling
  intervals in ads3/cft2.
\newblock {\em Journal of High Energy Physics}, 2021(12):1--31, 2021.

\bibitem{zhao2022charged}
Suting Zhao, Christian Northe, Konstantin Weisenberger, and Ren{\'e} Meyer.
\newblock Charged moments in w3 higher spin holography.
\newblock {\em Journal of High Energy Physics}, 2022(5):1--28, 2022.

\bibitem{Furukawa:2008uk}
Shunsuke Furukawa, Vincent Pasquier, and Jun'ichi Shiraishi.
\newblock {Mutual Information and Compactification Radius in a c=1 Critical
  Phase in One Dimension}.
\newblock {\em Phys. Rev. Lett.}, 102:170602, 2009.

\bibitem{Calabrese:2009ez}
Pasquale Calabrese, John Cardy, and Erik Tonni.
\newblock {Entanglement entropy of two disjoint intervals in conformal field
  theory}.
\newblock {\em J. Stat. Mech.}, 0911:P11001, 2009.

\bibitem{Calabrese:2010he}
Pasquale Calabrese, John Cardy, and Erik Tonni.
\newblock {Entanglement entropy of two disjoint intervals in conformal field
  theory II}.
\newblock {\em J. Stat. Mech.}, 1101:P01021, 2011.

\bibitem{Coser:2013qda}
Andrea Coser, Luca Tagliacozzo, and Erik Tonni.
\newblock {On R\'enyi entropies of disjoint intervals in conformal field
  theory}.
\newblock {\em J. Stat. Mech.}, 1401:P01008, 2014.

\bibitem{Headrick:2012fk}
Matthew Headrick, Albion Lawrence, and Matthew Roberts.
\newblock {Bose-Fermi duality and entanglement entropies}.
\newblock {\em J. Stat. Mech.}, 1302:P02022, 2013.

\bibitem{Coser:2015dvp}
Andrea Coser, Erik Tonni, and Pasquale Calabrese.
\newblock {Spin structures and entanglement of two disjoint intervals in
  conformal field theories}.
\newblock {\em J. Stat. Mech.}, 1605(5):053109, 2016.

\bibitem{Alba:2011fu}
Vincenzo Alba, Luca Tagliacozzo, and Pasquale Calabrese.
\newblock {Entanglement entropy of two disjoint intervals in c=1 theories}.
\newblock {\em J. Stat. Mech.}, 1106:P06012, 2011.

\bibitem{Rajabpour:2011pt}
M.~A. Rajabpour and F.~Gliozzi.
\newblock {Entanglement Entropy of Two Disjoint Intervals from Fusion Algebra
  of Twist Fields}.
\newblock {\em J. Stat. Mech.}, 1202:P02016, 2012.

\bibitem{Ruggiero:2018hyl}
Paola Ruggiero, Erik Tonni, and Pasquale Calabrese.
\newblock {Entanglement entropy of two disjoint intervals and the recursion
  formula for conformal blocks}.
\newblock {\em J. Stat. Mech.}, 1811(11):113101, 2018.

\bibitem{Chen:2013kpa}
Bin Chen and Jia-Ju Zhang.
\newblock {On short interval expansion of R\'enyi entropy}.
\newblock {\em JHEP}, 11:164, 2013.

\bibitem{Li:2016pwu}
Zhibin Li and Jia-ju Zhang.
\newblock {On one-loop entanglement entropy of two short intervals from OPE of
  twist operators}.
\newblock {\em JHEP}, 05:130, 2016.

\bibitem{Ares:2022gjb}
Filiberto Ares, Pasquale Calabrese, Giuseppe Di~Giulio, and Sara Murciano.
\newblock {Multi-charged moments of two intervals in conformal field theory}.
\newblock {\em JHEP}, 09:051, 2022.

\bibitem{Gaur:2023yru}
Himanshu Gaur and Urjit~A. Yajnik.
\newblock {Multi-charged moments and symmetry-resolved R\'enyi entropy of free
  compact boson for multiple disjoint intervals}.
\newblock {\em JHEP}, 01:042, 2024.

\bibitem{Calabrese:2012ew}
Pasquale Calabrese, John Cardy, and Erik Tonni.
\newblock {Entanglement negativity in quantum field theory}.
\newblock {\em Phys. Rev. Lett.}, 109:130502, 2012.

\bibitem{Calabrese:2012nk}
Pasquale Calabrese, John Cardy, and Erik Tonni.
\newblock {Entanglement negativity in extended systems: A field theoretical
  approach}.
\newblock {\em J. Stat. Mech.}, 1302:P02008, 2013.

\bibitem{Coser:2015eba}
Andrea Coser, Erik Tonni, and Pasquale Calabrese.
\newblock {Towards the entanglement negativity of two disjoint intervals for a
  one dimensional free fermion}.
\newblock {\em J. Stat. Mech.}, 1603(3):033116, 2016.

\bibitem{DeNobili:2015dla}
Cristiano De~Nobili, Andrea Coser, and Erik Tonni.
\newblock {Entanglement entropy and negativity of disjoint intervals in CFT:
  Some numerical extrapolations}.
\newblock {\em J. Stat. Mech.}, 1506(6):P06021, 2015.

\bibitem{Chen:2021nma}
Hui-Huang Chen.
\newblock {Charged R\'enyi negativity of massless free bosons}.
\newblock {\em JHEP}, 02:117, 2022.

\bibitem{Gaur:2022sjf}
Himanshu Gaur and Urjit~A. Yajnik.
\newblock {Charge imbalance resolved R\'enyi negativity for free compact boson:
  Two disjoint interval case}.
\newblock {\em JHEP}, 02:118, 2023.

\bibitem{Yin:2023jad}
Chao Yin and Shang Liu.
\newblock {Mixed-state entanglement measures in topological order}.
\newblock {\em Phys. Rev. B}, 108(3):035152, 2023.

\bibitem{Bruno:2023tez}
Andrea Bruno, Filiberto Ares, Sara Murciano, and Pasquale Calabrese.
\newblock {Symmetry resolution of the computable cross-norm negativity of two
  disjoint intervals in the massless Dirac field theory}.
\newblock {\em JHEP}, 02:009, 2024.

\bibitem{Giamarchi:2003ooa}
Thierry Giamarchi.
\newblock {\em {Quantum Physics in One Dimension}}.
\newblock Oxford University Press, 12 2003.

\bibitem{Dixon:1986qv}
Lance~J. Dixon, Daniel Friedan, Emil~J. Martinec, and Stephen~H. Shenker.
\newblock {The Conformal Field Theory of Orbifolds}.
\newblock {\em Nucl. Phys. B}, 282:13--73, 1987.

\bibitem{Verlinde:1986kw}
Erik~P. Verlinde and Herman~L. Verlinde.
\newblock {Chiral Bosonization, Determinants and the String Partition
  Function}.
\newblock {\em Nucl. Phys. B}, 288:357, 1987.

\bibitem{Dijkgraaf:1987vp}
Robbert Dijkgraaf, Erik~P. Verlinde, and Herman~L. Verlinde.
\newblock {C = 1 Conformal Field Theories on Riemann Surfaces}.
\newblock {\em Commun. Math. Phys.}, 115:649--690, 1988.

\bibitem{Alvarez-Gaume:1986rcs}
Luis Alvarez-Gaume, Gregory~W. Moore, and Cumrun Vafa.
\newblock {Theta Functions, Modular Invariance and Strings}.
\newblock {\em Commun. Math. Phys.}, 106:1--40, 1986.

\bibitem{Zamolodchikov:1987ae}
A.~B. Zamolodchikov.
\newblock {Conformal Scalar Field on the Hyperelliptic Curve and Critical
  Ashkin-teller Multipoint Correlation Functions}.
\newblock {\em Nucl. Phys. B}, 285:481--503, 1987.

\bibitem{mumford2007tata}
David Mumford, Madhav Nori, and Peter Norman.
\newblock {\em Tata lectures on theta III}, volume~43.
\newblock Springer, 2007.

\bibitem{Miki:1987mp}
Kei Miki.
\newblock {Vacuum Amplitudes Without Twist Fields for $Z(N$) Orbifold and
  Correlation Functions of Twist Fields for $Z$(2) Orbifold}.
\newblock {\em Phys. Lett. B}, 191:127--134, 1987.

\bibitem{rauch1970theta}
Harry~E Rauch and Hershel~M Farkas.
\newblock Theta constants of two kinds on a compact riemann surface of genus 2.
\newblock {\em Journal d’Analyse Math{\'e}matique}, 23(1):381--407, 1970.

\bibitem{farkas1970period}
Hershel~M Farkas and Harry~E Rauch.
\newblock Period relations of schottky type on riemann surfaces.
\newblock {\em Annals of Mathematics}, 92(3):434--461, 1970.

\bibitem{fay2006theta}
John~David Fay.
\newblock {\em Theta functions on Riemann surfaces}, volume 352.
\newblock Springer, 2006.

\bibitem{Alvarez-Gaume:1987xem}
Luis Alvarez-Gaume, C.~Gomez, and C.~Reina.
\newblock {NEW METHODS IN STRING THEORY}.
\newblock 6 1987.

\bibitem{Alvarez-Gaume:1986bwm}
Luis Alvarez-Gaume and Philip~C. Nelson.
\newblock {RIEMANN SURFACES AND STRING THEORIES}.
\newblock In {\em {4th Trieste Spring School on Supersymmetry, Supergravity,
  Superstrings}: {(followed by 3 day Workshop)}}, 12 1986.

\bibitem{Bueno:2020vnx}
Pablo Bueno and Horacio Casini.
\newblock {Reflected entropy, symmetries and free fermions}.
\newblock {\em JHEP}, 05:103, 2020.

\bibitem{Bueno:2020fle}
Pablo Bueno and Horacio Casini.
\newblock {Reflected entropy for free scalars}.
\newblock {\em JHEP}, 11:148, 2020.

\bibitem{Peschel:2002yqj}
Ingo Peschel.
\newblock {Calculation of reduced density matrices from correlation functions}.
\newblock {\em J. Phys. A}, 36(14):L205, 2003.

\end{thebibliography}
% BIBLIOGRAPHY
% use BIBTEX if you want
%\bibliographystyle{unsrt}
%\bibliography{Biblio1.bib}
% \bibitem{a}
% Author, \emph{Title}, \emph{J. Abbrev.} {\bf vol} (year) pg.

% \bibitem{b}
% Author, \emph{Title},
% arxiv:1234.5678.
% \bibitem{c}
% Author, \emph{Title},
% Publisher (year).

% Please avoid comments such as "For a review'', "For some examples",
% "and references therein" or move them in the text. In general,
% please leave only references in the bibliography and move all
% accessory text in footnotes.

% Also, please have only one work for each \bibitem.
\end{document}